\documentclass[twocolumn]{aastex631}

\received{November 15, 2022}
\revised{December 12, 2022}
\accepted{December 13, 2022}
\submitjournal{AJ}

\shorttitle{MeerKAT Holography Measurements in the \em UHF\em, \em L\em, and \em S \em bands}
\shortauthors{de Villiers}

\usepackage{amsmath}
\usepackage{multirow}
\graphicspath{{./}{figures/}}

\begin{document}

\title{MeerKAT Holography Measurements in the \em UHF\em, \em L\em, and \em S \em bands}

\correspondingauthor{Mattieu S. de Villiers}
\email{mattieu@sarao.ac.za}

\author[0000-0001-5628-0417]{Mattieu S. de Villiers}
\affiliation{South African Radio Astronomy Observatory,
2 Fir Street, Black River Park, Observatory, 7925, RSA}

\begin{abstract}

Radio holographic measurements using the MeerKAT telescope are presented for each of its supported observing bands, namely \em UHF \em (544--1087 MHz), \em L \em (856--1711 MHz) and \em S \em (1750--3499 MHz). Because the \em UHF\em-band receiver design is a scaled version of that of the \em L \em band, the electromagnetic performance in these two bands are expectedly similar to one another. Despite also being linearly polarized, \em S\em-band receivers have an entirely different design and distinct performance characteristics from the lower two bands. As introduced in previous work for the \em L \em band, evidence of higher-order waveguide mode activation also appears in \em S\em-band measurements but there are differences in its manifestation. Frequency-dependent pointing (beam squint), beam width, beam ellipticity, errorbeam, instrumental polarization and cross-polarization power measurements are illustrated for each of MeerKAT's observational bands in a side-by-side style to facilitate the comparison of features. The derivation of collimation errors and main reflector surface errors from measurements made at these relatively low observation frequencies is also discussed. Results include elevation and ambient temperature effects on collimation, as well as the signatures of collimation degrading over time. The accompanying data release includes a snapshot of full Jones matrix primary beam patterns for all bands and antennas, with corresponding derived metrics.

\end{abstract}


\section{Introduction} \label{sec:intro}

\begin{deluxetable*}{ccccccc}[t]
\tablenum{1}
\tablecaption{Overview of the half-array scan cycles captured for monitoring and characterization of MeerKAT primary beams, up until March 2022.\label{tab:measurements}}
\tablewidth{0pt}
\tablehead{
\colhead{Band} & \colhead{Inception date} &\colhead{Scan cycles} & \colhead{Scan extent} & \colhead{Scan duration} & \colhead{Sampling period} & \colhead{Storage} 
}
\startdata
\em UHF \em & July 2019 & 458 & 15$^\circ$ & 45 -- 90 min & 1 s & 125 TB\\
\em L \em & March 2019 & 887 & 10$^\circ$ & 30 -- 60 min & 1 s & 175 TB\\
\em S0 \em & February 2021 & 95 & 6$^\circ$ & 60 min & 2 s & 13 TB\\
\em S4 \em & February 2021 & 87 & 4$^\circ$ & 60 min & 2 s & 12 TB\\
\enddata
\end{deluxetable*}

Full-polarization \em L-\em band (856--1711 MHz) primary beam measurements of the 64 antenna (13.5 m diameter) MeerKAT radio synthesis telescope were recently published \citep{deVilliers_2022}. That precursory work details how per-receiver variability in the activation of higher-order waveguide modes studied by \cite{deVilliers_2021} affect frequency and polarization dependent pointing and shape characteristics in the upper half of this band for the antennas in the array. The importance of applying a reliable pointing error estimate, or invoking reference pointing, is also emphasized when using high-fidelity primary beam patterns.

The purpose of this follow-up paper is to share with the broader community the measured primary beam characteristics of the MeerKAT's remaining observable bands, namely \em UHF \em (544--1087 MHz) and \em S \em (1750--3499 MHz). Expectedly, the \em UHF-\em band performance is akin to that of the \em L \em band because their feeds and orthomode transducers (OMTs) share a common, scaled design. These electromagnetic (EM) components were designed and manufactured by EMSS in South Africa \citep{Lehmensiek_2012}. With distinctive performance characteristics, the \em S\em-band feeds and OMTs \citep{kramer_2016} were independently designed and supplied by the Max Planck Institute for Radio Astronomy (MPIfR) and were installed on MeerKAT more recently, during the course of 2021. 

Although each of the MeerKAT feed bands spans a full octave, the receivers are designed to meet specific performance criterion over $\sim 85$ \% of this range. The nuanced results illustrated in this paper fall within the design performance limits set for the telescope. 

Beam metrics are plotted alongside each other to reveal how waveguide mode features manifest in the primary beams differently across each band. While the errorbeam metric is used to quantify the variability in the beam shapes amongst different antennas and environmental conditions, the instrumental polarization metric is used to quantify the polarization purity performance over the mainlobe.

The scope of the results presented extends beyond the far-field pattern to include its Fourier transform, namely, the aperture plane distribution \citep{rochblatt2008}. From the aperture plane phase, collimation (reflector optics alignment) measurements are determined and provided for all the antennas. Trends over time and common dependencies on environmental conditions are quantified in the results. 

The accurate measurement of collimation errors for reflector antennas at microwave frequencies below 8 GHz is rare because of the increasing effects of multi-path interference and scatter \citep{Popping_2008} off feeds, sub-reflector extensions and other antenna structures. \cite{Perley_2016} also reports a stronger frequency dependence of the EVLA primary beam profiles in lower frequency bands. Nevertheless it will be shown that reliable collimation estimates can be achieved subject to the availability of wide bandwidth data and a large number of antennas in the array.

In addition to the mode activations that are observed in the \em UHF \em and \em L \em bands, there appears to be unexpected ghost modes \citep{Jaynes_1958} that propagate in the \em S \em band, which dramatically change the pointing direction of the beams at intermittent frequencies (see Section \ref{section:beam_center_and_squint}). Ghost modes, which are resonances below the cut-off frequency of the corresponding activated mode, are usually confined to the vicinity of imperfections in a transmission line waveguide. In the context of antennas however, given the proximity of the feed horn to small scale design features of the waveguide, these modes can propagate \citep{Morgan_2008}. This effect has been researched extensively for the ALMA telescope \citep{Gonzales_2014} when it was discovered that the design specifications were not met at some frequencies. Beam elongation effects that beset the instrumental polarization may pose further challenges in observations above 3 GHz using the MeerKAT \em S\em -band receivers.

The next section discusses the scope of the data that has been acquired, mentions reduction tools used, and explains how measurements are combined to suppress radio frequency interference (RFI). Section \ref{section:primary_beam_characteristics} compares the measured far-field beam plane characteristics amongst the \em UHF\em, \em L\em , and \em S \em bands, while Section \ref{section:aperture_plane_results} elaborates on the aperture plane derived results, featuring collimation error details.

\begin{deluxetable*}{cccccccccc}[t]
\tablenum{2}
\tablecaption{Average S/N and maximum elevations recorded during measurements of prevailing sources observed over the indicated bands. Flux density information of the sources are quoted from the online ATCA Calibrator Database for a survey in 2020 (except for PKS 0408-65 and 3C 454.3 dated 2011 and 2000, respectively). \label{tab:targets}}
\tablewidth{0pt}
\tablehead{
\colhead{Source name} & \multicolumn3c{ATCA flux density (Jy)}& \colhead{Max elevation} & \colhead{\em UHF \em} &\colhead{\em L \em} & \colhead{\em S0 \em} & \colhead{\em S4 \em} \\
\colhead{}  & \colhead{816 MHz} & \colhead{1284 MHz} & \colhead{2624 MHz} & \colhead{} & \colhead{(S/N)} &\colhead{(S/N)} & \colhead{(S/N)} & \colhead{(S/N)} 
}
\startdata
PKS 0408-65 & 29.3 & 17.1 & 7.5 & 55$^\circ$& 86 & 66 & 33 & 24\\
3C 279 & 11.7 & 10.0 & 10.1 & 65$^\circ$& 67 & 53 & 36 & 39\\
PKS 1934-63 & 14.4 & 15.2 & 10.7 & 57$^\circ$& 64 & 63 & &\\
PKS 0023-26 & 12.1 & 9.2 & 5.8 & 83$^\circ$& 61 & 49 & 26 & 24\\
J1924-2914 & 5.4 & 5.0 & 5.7 & 80$^\circ$& 30 & 30 & 18 & 32\\
J0825-5010 & 4.6 & 5.9& 5.6 & 70$^\circ$& 26 & 32 & 22 & 25\\
3C 454.3 & 16.4 & 14.9 & 12.7 & 42$^\circ$& & & 50 & 63\\
PKS 1421-490 & 10.0 & 8.7 & 6.8 & 70$^\circ$& & & 26 & 27\\
\enddata
\end{deluxetable*}

\section{The captured data}

Although routine holographic monitoring of primary beams and collimation errors commenced in November 2017 for the \em L \em band, datasets prior to March 2019 are excluded from this study. The storage of MeerKAT's visibility data transitioned from a self-contained hierarchical data format (hdf5) to a distributed cluster archive system (Ceph) which required new data access patterns for efficient processing.

Data capture in the \em UHF \em and much later in the \em S \em bands started during their respective roll-out phases as more antennas became available with these receivers installed. At first the effort was driven by the need for formal qualification and acceptance, which later extended into monitoring for more thorough antenna characterization. Holography observations are usually scheduled alongside other telescope operations activities when insufficient antennas are available for science observations due to antenna maintenance. In practice the holography observations are quick and robust, and well-suited to use up any idle time.

Table \ref{tab:measurements} summarizes the scope of the datasets captured in the various bands that is reported on in this work. As described more comprehensively in \cite{deVilliers_2022}, during each half-array scan cycle only half of the available antennas are scanned while the other half tracks the target. The scan extent parameters of the beam measurements are selected to include approximately two sidelobes at the lowest and five at the highest frequency in each band. As such, the scan extents range from 4$^\circ$ in the \em S \em to 15$^\circ$ in the \em UHF \em band. 

In the \em L \em band, a scan cycle usually takes 30 minutes to observe, but at elevations higher than 60$^\circ$, to avoid straining the antenna drive system, the duration is increased accommodating the longer azimuthal distances that need to be travelled. Correspondingly, for the \em UHF \em band the minimum scan duration is 45 minutes. Perhaps counterintuitively, the scan-cycle duration is increased to a fixed hour in the \em S \em band due to the weaker flux density of available celestial targets, and lower sensitivity, for this band. The sampling period is doubled to two seconds to keep data volumes and processing speed comparable to the other bands.

While the unprocessed visibilities of a typical 30 minute \em L\em-band scan cycle using 61 (out of 64) antennas and 1024 channels occupy nearly 200 GB of archive space, only about 1.3 GB is needed to store its full aperture plane reduction result. The archive only keeps newly captured visibility data accessible on disk for about 6 months, after which it can be restaged from tape upon request. The vast majority of measurements are recorded in the \em L \em band (887 scan cycles) while only about 12 \% of the measurements are in the \em S \em band.

Unlike for the \em UHF \em and \em L \em bands, the \em S \em band is divided into 5 overlapping subbands due to its wide extent. The current MeerKAT correlator can process only one of these subbands at a time. The lowest subband, \em S0\em, borders with the highest, \em S4\em, at 2625 MHz. Overlooking bandpass roll-off challenges, together these two outer subbands cover the whole of the observable \em S \em band. In terms of holography it is sufficient to observe in these two subbands only to characterize the entire \em S \em band. Overlapping subbands are useful for continuity checks.

MeerKAT feed bands were designed to perform over a narrower range than the full bandwidth that is captured (\em UHF\em: 580--1015 MHz \em L\em: 900--1670 MHz, \em S\em: unspecified). Outside these ranges, attenuation and electromagnetic effects on the passband may undermine the scientific quality of the data as a consequence of the design stage trade-offs that were permitted.

Table \ref{tab:targets} lists the most promising celestial radio sources that were observed in this study and quotes the average S/Ns that were achieved in their measurements for the respective bands. The maximum elevations are also included to highlight the fact that high elevation data is only obtainable at limited opportunities, and using lower flux sources. The observation script prioritizes available sources based on their elevation, flux density and angular distance from the Sun with the goal to add desirable measurements that best complement those that have already been captured. Flux density information of the sources as extracted from the online ATCA Calibrator Database\footnote{\href{https://www.narrabri.atnf.csiro.au/calibrators/}{https://www.narrabri.atnf.csiro.au/calibrators/}} is listed for comparison and shows that the \em S-\em band targets have lower fluxes.

The MeerKAT holography framework was used to observe and process the beam measurements into results that are stored compactly as aperture plane data cubes alongside derived performance metrics in a database for all antennas per band. The databases are accessible programmatically using features that can select, model, display and report on various measurement combinations and products of interest. Beam patterns are reconstructed on demand by means of the inverse discrete Fourier transform implemented on GPUs. The results reported in the upcoming sections were extracted from these databases.

Remarkably, the long baseline antenna m060 is known to experience almost no terrestrial RFI in the \em UHF \em band because of the unique way it is nestled in the Karoo landscape. Likewise, because satellites move across sidelobes while the celestial sources are observed, the impact of RFI is varied across different datasets and its contribution is random.

The success of eliminating RFI by combining data from several observations is much attributed to the weighting employed because impaired data need to be downweighted appropriately. A single measurement using $N$ tracking antennas yields $N$ results for each of the scanning antennas. Measurement noise, including much of the RFI is independent across these results so that, for each of the scanning antennas the $N$ results are combined into a weighted average. In this work the weighting, $1/\sigma^2_\mathrm{scan,track}$ below, is determined from the standard deviation $\sigma_\mathrm{scan,track}(f)$ of the aperture plane noise that is measured outside the physical aperture for individual scanning to tracking antenna baselines:
\begin{equation}
\frac{1}{\sigma^2_\mathrm{scan}(f)}=\sum_\mathrm{track}{\frac{1}{\sigma^2_\mathrm{scan,track}(f)}}
\end{equation}
\begin{equation}
A_\mathrm{scan}(x,y,f)=  \sigma^2_\mathrm{scan}(f) \sum_\mathrm{track}{\frac{A_\mathrm{scan,track}(x,y,f)}{\sigma^2_\mathrm{scan,track}(f)}}
\end{equation}
where $A_\mathrm{scan,track}(x,y,f)$ refers to a complex aperture plane measurement from the baseline between a particular scanning and tracking antenna pair. In this paper the $y$ axis is aligned parallel to the vertical or elevation plane of symmetry, while the $x$ axis is parallel to the horizontal or cross-elevation plane. Once pointing errors ($\sigma \approx$ 0.6 arcminutes) are removed, measurements captured under comparable environmental conditions yet at different times and of varying S/N quality, can also constructively be combined by such a weighted sum, for each scanning antenna independently. 

\begin{figure*}[t]
\centering
\includegraphics[width=\linewidth,trim=2cm 1.5cm 3cm 2cm, clip]{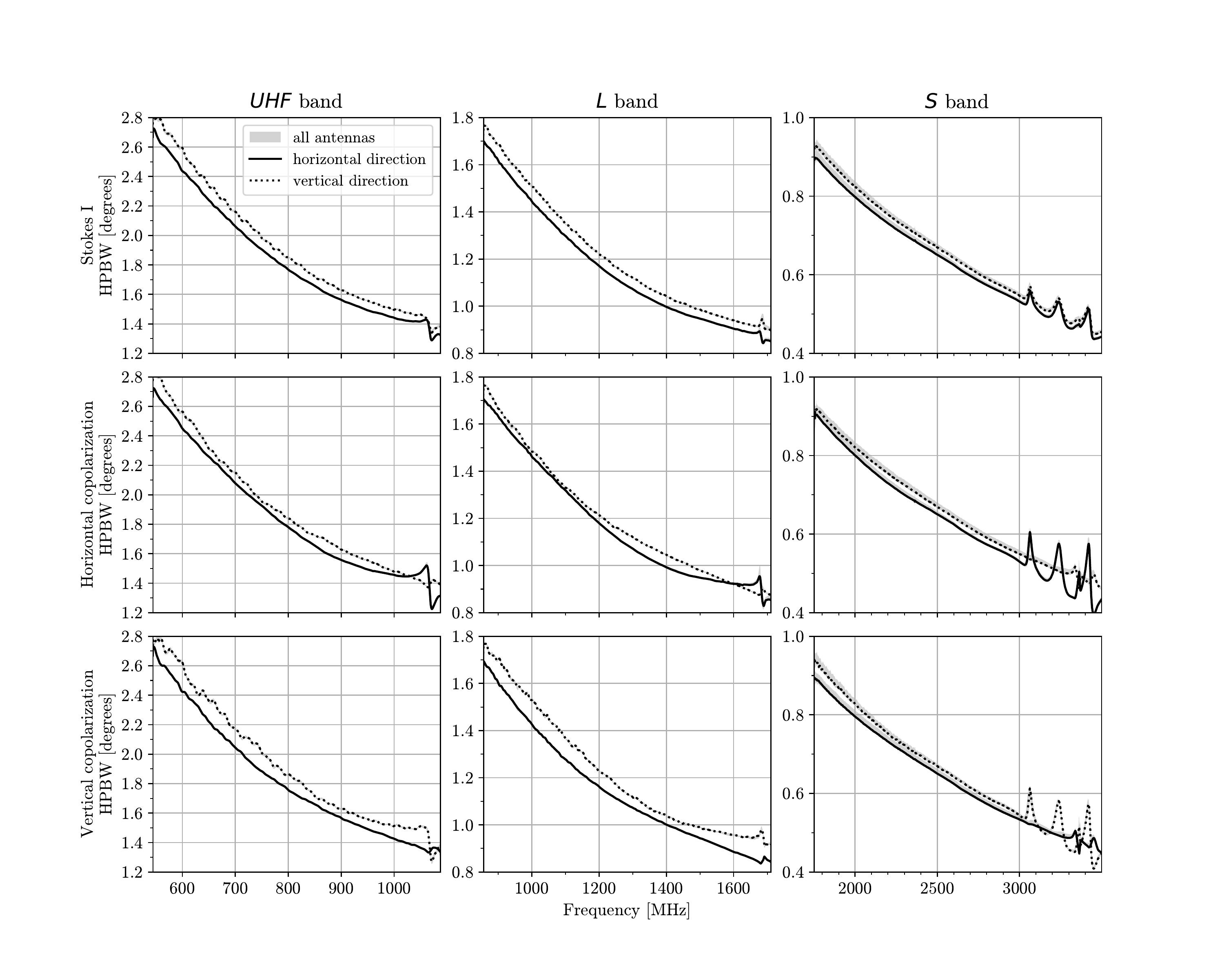}
\caption{The half-power beam width measurements show that the MeerKAT Stokes I primary beams are consistently narrower in the horizontal direction across all bands. Very narrow margins of variation in beam widths are noted amongst all antennas in the \em UHF \em and \em L \em bands. Dramatic beam shape elongation features occur as a result of the activation of the TM11 waveguide mode towards the upper end of each band due to respective OMT designs, more prominently in the \em S \em band. 
\label{fig:hpbw}}
\end{figure*}

\section{Primary Beam Characteristics}
\label{section:primary_beam_characteristics}

In this section, the most prominent qualities of the measured primary beams for all of MeerKAT's operational observing bands are compared against each other. All results provided below (except where explicitly stated otherwise in Section \ref{section:perspective_on_beam_shape_variability} and Section \ref{section:aperture_plane_results}) are referred to a 60$^\circ$ antenna elevation (reflector collimation was tuned using photogrammetry during assembly at 61$^\circ$, the \em rigging angle\em) and 15$^\circ$C ambient temperature, averaged over the time period from January 2020 to January 2022. However, the \em S-\em band results are averaged likewise from February 2021 to February 2022. The measured properties and corresponding gridded beam cubes for all bands will be made available electronically\footnote{\href{https://doi.org/10.48479/wdb0-h061}{doi: 10.48479/wdb0-h061}} for the benefit of the reader.

\subsection{Beam width, ellipticity and orientation}

\begin{figure*}[t]
\centering
\includegraphics[width=\linewidth,trim=2cm 1.5cm 3cm 2cm, clip]{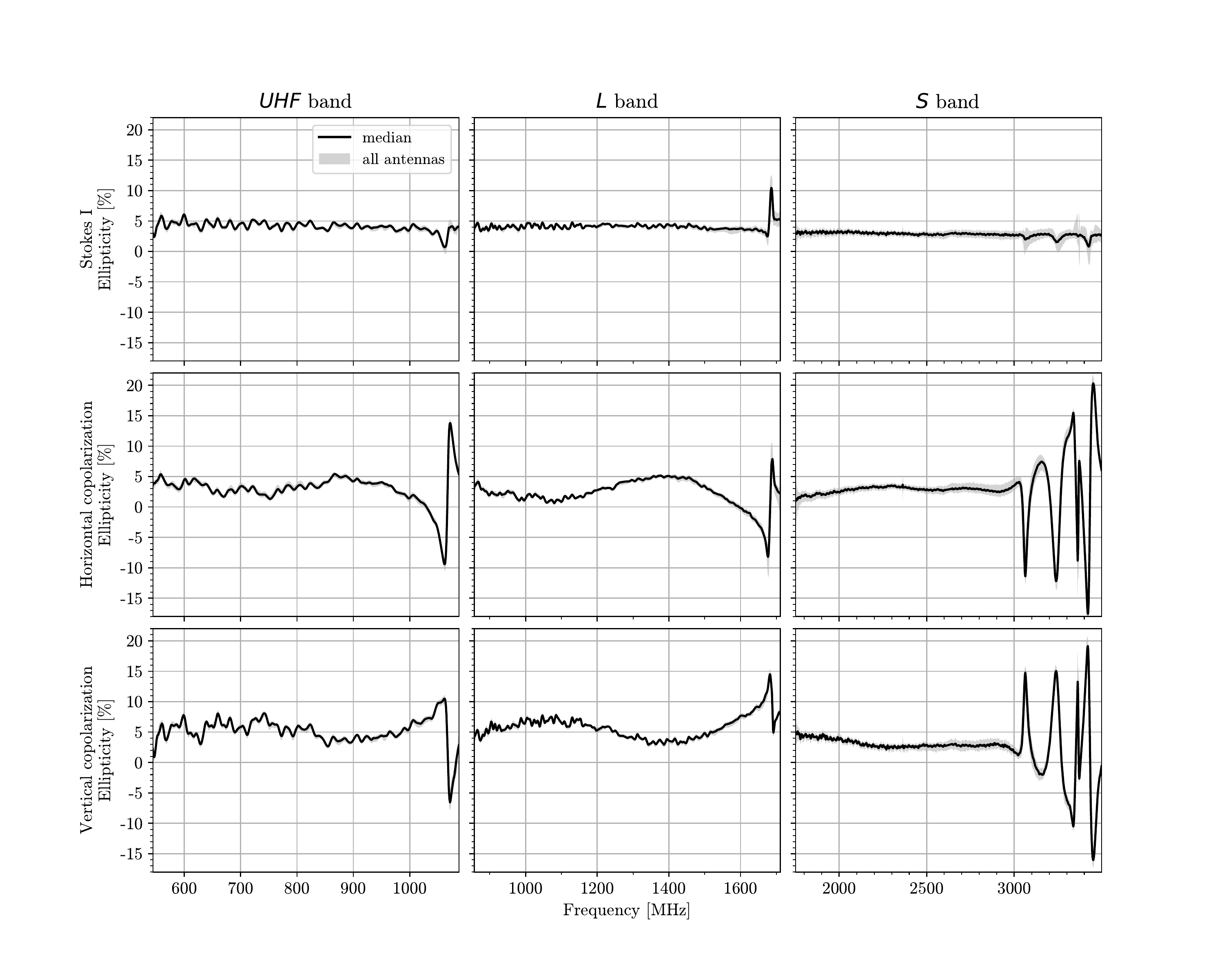}
\caption{Primary beam ellipticity as a function of frequency illustrates the activation of the TM11 mode towards the right of each respective band where the copolarization beams are complementarily elongated. A larger fraction of the observable band is severely affected by this mode in the \em S \em band, beyond 3 GHz. There remains a narrow margin of variation in beam ellipticity amongst the different antennas in all bands, with somewhat more variability in the \em S \em band.
\label{fig:ellipticity}}
\end{figure*}

While the Stokes I primary beam width is proportional to the usual $\lambda \slash D$ ratio over most of each band, this approximation is less true for the individual copolarization (i.e. parallel hand) beams, as illustrated in Figure \ref{fig:hpbw}. Except at limited frequencies, all MeerKAT primary beams are elongated in the vertical direction because the antenna main reflector design is wider in the horizontal direction as projected onto the aperture plane. Towards the upper end of each band, and more impressively in the \em S \em band above 3 GHz, the beam width expands and contracts as a function of frequency due to the activation of the TM11 mode in the waveguide of the OMT that house the dipoles of the feed. It appears for the \em S \em band that the operational frequency range starts relatively further up above the lower cut-off of the fundamental mode than for the \em UHF \em and \em L \em bands, so that the available band exposes more higher order modes. For the lower frequency bands, similar features coincide with the band select filter gain falloff, which deters users from utilizing data captured outside the band specification limits.

The percentage by which the beams are elongated is more clearly depicted in Figure \ref{fig:ellipticity}. While the \em UHF\em- and \em L\em-band Stokes I primary beams have ellipticities approaching 5\%, the ellipticity only reaches about 3\% in the \em S \em band.

\begin{figure*}[t]
\centering
\includegraphics[width=\linewidth,trim=2cm 1.5cm 3cm 2cm, clip]{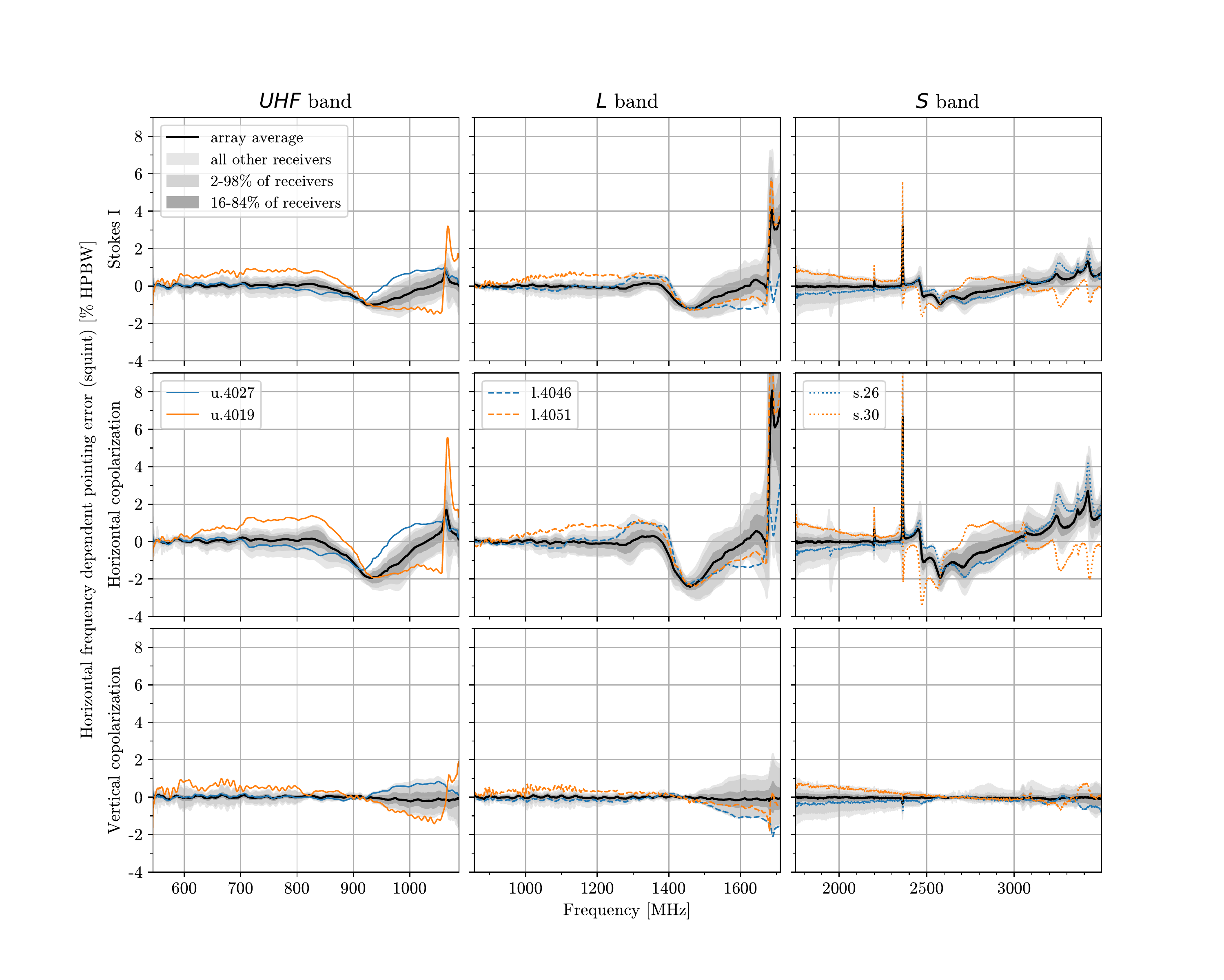}
\caption{Frequency-dependent pointing of MeerKAT primary beams in the horizontal direction reveals the activation of the asymmetrical TE21 waveguide mode as deviations to the left and right. Results for a few outlier receivers are indicated using color to emphasize that, even though a range of variation is indicated for many different receivers in greyscale, the characteristics of individual receivers are unique and do not change over time. Antenna pointing in the horizontal direction is zeroed at 900, 1420 and 2625 MHz in the \em UHF\em , \em L \em  and \em S \em  bands respectively, using the vertical copolarization profile. Note that the overlapping 856--1088 MHz range of the \em L \em band receiver may be preferred in polarization studies over that of the \em UHF \em receiver due to the increasing effect of higher order waveguide modes towards the upper part of each band.
\label{fig:pointingH}}
\end{figure*}

Despite the small range in variability seen in the beam width and elongation plots amongst antennas, substantial differences in beam shape may still occur due to the orientation in which the beam power patterns are elongated. A skew effecting the primary beam shape happens due to errors in the feed indexer offsets (see later in Figure \ref{fig:collimation_date}) amidst the offset Gregorian antenna design. An extreme of 25$^\circ$ apparent rotation of the best fit ellipse through the Stokes I power beam was recorded for antenna m054 in late June 2021 which was later conclusively attributed to a 4.2$^\circ$ feed indexer error. The rotational effect on the copolarization power beam is in fact a shear deformation that should not be confused with the effect of the dipole orientation which acts independently and is instead subject to assembly and manufacturing tolerances of the OMT that rotates the feed pattern axially. The measured dipole orientation variability in each band can be summarized as \em UHF\em:  $\sigma \approx 0.8^\circ$, \em L\em: $\sigma \approx 0.7^\circ$ and \em S\em: $\sigma \approx 1^\circ$. While larger orientation errors are more common in the \em S \em band, the receiver l.4002 installed on m006, since August 2019, has a $\sim 3^\circ$ error that is exceptional for \em L \em band. Jointly the beam shear and dipole orientation influence the instrumental polarization response for an antenna.

\subsection{Beam center and squint behaviour}
\label{section:beam_center_and_squint}
\begin{figure*}[t]
\centering
\includegraphics[width=\linewidth,trim=2cm 1.5cm 3cm 2cm, clip]{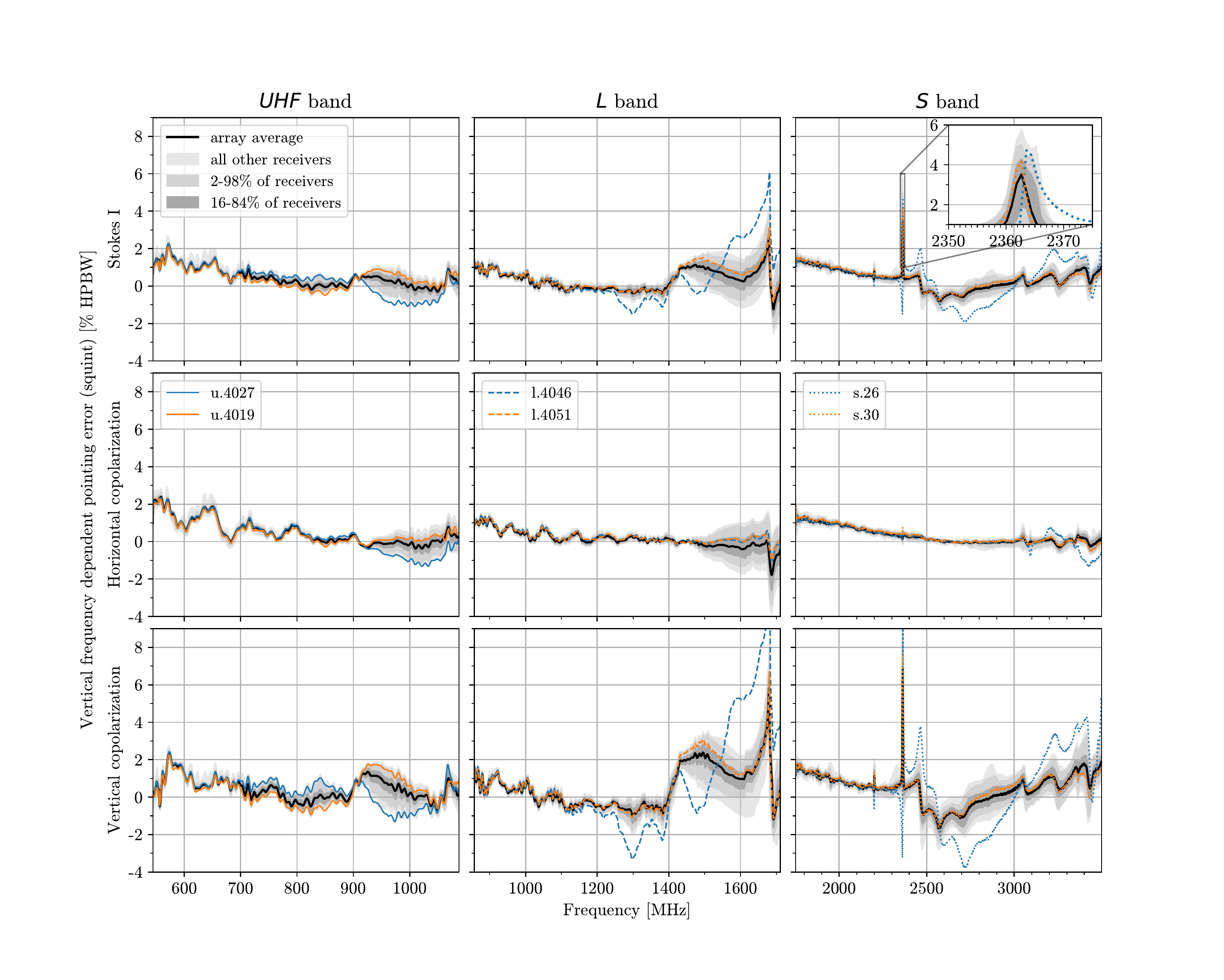}
\caption{Frequency-dependent pointing of MeerKAT primary beams in the vertical direction reveals the activation of the asymmetrical TE21 waveguide mode as deviations up and down. Antenna pointing in the vertical direction is zeroed at 900, 1420 and 2625 MHz in the \em UHF\em , \em L \em  and \em S \em  bands respectively, using the horizontal copolarization profile. An inset axis highlights the resonant activation of a ghost mode (of TE21) that propagates, and differs in character slightly per receiver.
\label{fig:pointingV}}
\end{figure*}

As shown in Figures \ref{fig:pointingH} and \ref{fig:pointingV} for the horizontal and vertical directions respectively, the best fit centroid of the primary beams varies systematically as a function of frequency over each band by at least 2\% of the half-power beam width, and more conspicuously, by at least 6\%, at a few frequencies in the \em S \em band. This frequency dependent pointing property of the beam might equally well be considered by some as a change in beam shape rather than position.

Frequency dependent pointing characteristics differs for each copolarization, resulting in frequency dependent squint between the parallel hand beams. 
As seen in the third row of Figure \ref{fig:pointingH}, the vertical copolarization beam varies the least in the horizontal direction while this beam meanders vertically (third row of Figure \ref{fig:pointingV}) as a function of frequency. Conversely, the horizontal copolarization beam varies the least over frequency in the vertical direction. Nevertheless, due to the offset-Gregorian reflector optics design, an intrinsic profile is sustained in the vertical direction that is more obtrusive at lower frequencies in each band (as seen in row 2 of Figure \ref{fig:pointingV}). Due to the meandering centroid behaviour, it is sensible to pin antenna pointing to the horizontal and vertical offsets of the vertical and horizontal copolarization beams respectively, instead of the Stokes I beam.

A large range of variation amongst different antennas is observed in terms of its frequency dependent pointing. Some outlier cases are striking, and are highlighted in Figures \ref{fig:pointingH} and \ref{fig:pointingV} using color. The frequency dependent pointing in the upper half of each band is mostly attributed to the activation of the TE21 mode in the waveguide of the OMTs, and these are sensitive to manufacturing tolerances that results in significant receiver-specific differences.

The varied sloped squint errors for outlier antennas noticed in the \em S0 \em band increasingly towards lower frequency in Figure \ref{fig:pointingH} is caused by feed position errors in the horizontal direction due to the feed indexer of affected antennas.

It may be expected that it is more accurate to determine pointing at the highest frequency because the beam is naturally narrower; however, such a strategy may not be effective in this case because the squint characteristics vary more here due to sensitivity to manufacturing tolerances.

\begin{figure*}[t]
\centering
\includegraphics[width=\linewidth,trim=2cm 2cm 3cm 2.1cm, clip]{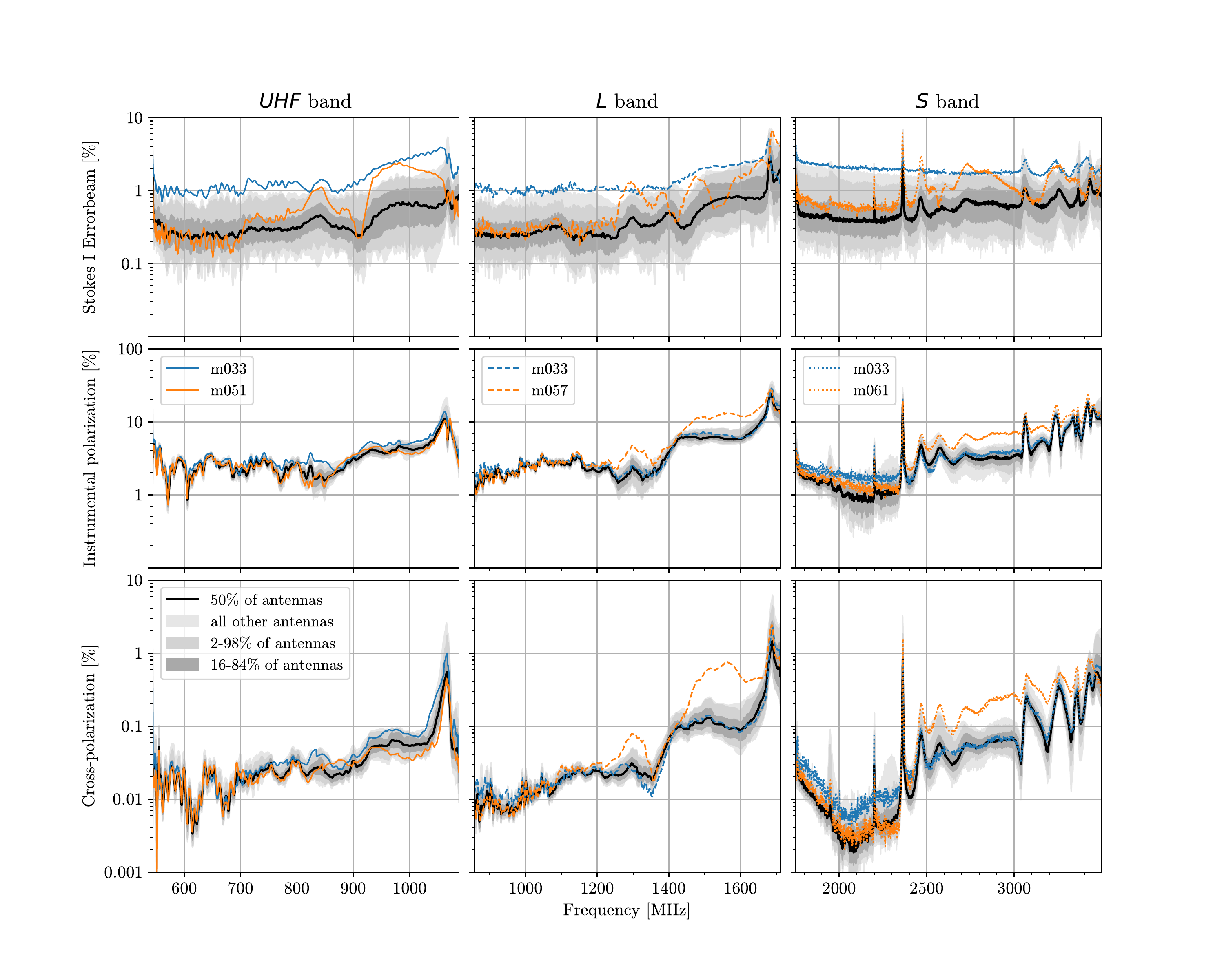}
\caption{The maximum difference in directivity power between the Stokes I beams for all antennas and that of the array average is shown in the top row of panels for the respective bands as evaluated over the half-power region of the beam after accounting for antenna pointing errors. These errorbeam performance profiles show that larger variations of the beam shape amongst antennas exist in the lower half of the \em S \em band compared to that of the other bands. In the second row the maximum instrumental polarization induced over the half-power region of the beam is shown (not \em instrumental polarization error \em as in Table \ref{tab:errorbeam}). While the instrumental polarization seems improved in the \em S \em band at more frequencies than the other bands, some more spikes of poorer performance occur at intermittent frequencies. The last row shows the maximum power of either cross-polarization beam that occurs over the half-power region of the Stokes I beam relative to the peak Stokes I power. A few outlier cases are illustrated in color. The degraded Stokes I errorbeam performance of m033 in the top row is mostly affected by an exceptionally large feed indexer error which manifests as a horizontal (cross-elevation) feed position error causing beam shear and coma lobes. It is interesting to note that despite m033's poor Stokes I performance, its polarization performance is not as bad because the cross-polarization pattern is more (but not only) sensitive to the characteristics of the feed package installed rather than reflector collimation effects.
\label{fig:errorbeam_uls}}
\end{figure*}

\subsection{Errorbeam}
\label{section:errorbeam}
The errorbeam metric \citep{deVilliers_2022} is used to quantify the maximum power by which the beam shape of one antenna differs from a reference pattern. In this context the reference is the average beam shape at 60$^\circ$ elevation. The design criteria allowed a maximum error of 4\% at any frequency over the usable range of the band in question in order to pass qualification. While the errorbeam variations amongst antennas in the upper half of each band is mostly attributed to a particular OMT installed, the poor performance of a few antennas across all bands is caused by sizable feed indexer offsets, notably in the case of antenna m033. 

\begin{figure*}[t]
\centering
\includegraphics[width=\linewidth,trim=2cm 1cm 3cm 1.5cm, clip]{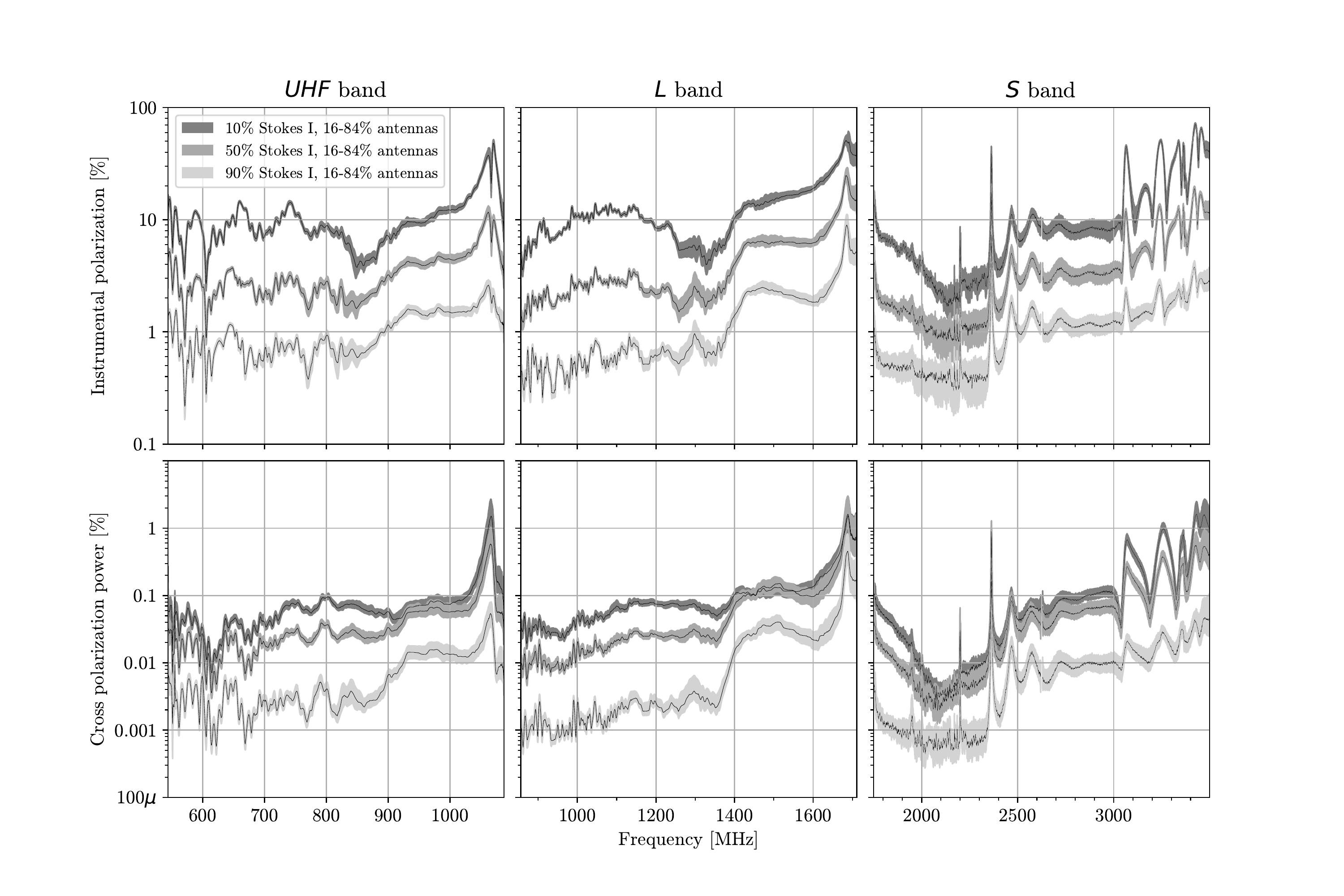}
\caption{Maximum instrumental polarization and cross-polarization power as a function of frequency is shown at 3 different contour levels of Stokes I power of the mainlobe. The envelopes at each power level show the spread over 16-84\% of antennas, representative of one standard deviation of variation amongst the different antennas.
\label{fig:instrumental_cross_intensity}}
\end{figure*}

The first row of Figure \ref{fig:errorbeam_uls} shows statistical errorbeam results of the Stokes I beam shape for the different bands. Two outlier antennas for each band are illustrated in color to emphasize the fact that individual antennas have specific performance that does not change unless a physical change is made, such as swapping out a receiver or if a collimation adjustment is made. Percentile ranges are shown such that, for example, the dark grey band reveals the range of variation in errorbeam for 16--84\% of the antennas. It is noteworthy that the range of variation in errorbeam (and hence differences in beam shape relative to the reference) is larger in the upper part of each band compared to the lower half. This is largely because beam squint relative to that of the reference pattern is counted towards pattern distortion.

\subsection{Instrumental polarization}

Instrumental polarization quantifies how much spurious Stokes Q,U,V power in total is inadvertently induced off boresight due to instrumental effects when observing an unpolarized point source target. As such, it conveys a more intuitive interpretation of the polarization performance of the primary beam to scientists than cross-polarization beam patterns. Instrumental polarization, $P$ below, is a function of both the co- and cross-polarization patterns:
\begin{equation}
P=\sqrt{Q^2+U^2+V^2} \slash I
\end{equation}
where the Stokes power beams are
\begin{eqnarray}
I&=&\frac{1}{2}\left(|E_{\mathrm{HH}}|^2 +|E_{\mathrm{HV}}|^2 +|E_{\mathrm{VH}}|^2 +|E_{\mathrm{VV}}|^2\right)\\
Q&=&\frac{1}{2}\left(|E_{\mathrm{HH}}|^2 +|E_{\mathrm{HV}}|^2 -|E_{\mathrm{VH}}|^2 -|E_{\mathrm{VV}}|^2\right)\\
U&=&\Re\left(E_{\mathrm{HH}}\overline{E_{\mathrm{VH}}} +E_{\mathrm{HV}}\overline{E_{\mathrm{VV}}}\right)\\
V&=&\Im\left(E_{\mathrm{HH}}\overline{E_{\mathrm{VH}}} +E_{\mathrm{HV}}\overline{E_{\mathrm{VV}}}\right)
\label{eqn:stokeserrorbeam}
\end{eqnarray}
and the horizontal and vertical linearly polarized copolarization voltage beams are $E_{\mathrm{HH}}$ and $E_{\mathrm{VV}}$ respectively, and the linear cross-polarization voltage beams are $E_{\mathrm{HV}}$ and $E_{\mathrm{VH}}$.

Complementary to the errorbeam metric, it is convenient to define the \em instrumental polarization error \em metric as the maximum absolute difference in instrumental polarization that occur over the half-power region of the beam when comparing two beams. This metric summarizes by how much the polarization properties of two beams differ from each other as a function of frequency, and is quantified in units of percentage power. Section \ref{section:perspective_on_beam_shape_variability} shows by how much the instrumental polarization changes for a few common scenarios using this metric.

\begin{deluxetable*}{@{\extracolsep{4pt}}ccccccc}[!t]
\tablenum{3}
\tablecaption{A summary of errorbeam and instrumental polarization error results. In each case, a different comparand is assessed against the time-averaged per-antenna reference primary beam pattern measurements at 60$^\circ$ elevation and 15$^\circ$C. The environmental conditions remain the same for the comparand except for aspects explicitly indicated otherwise. \label{tab:errorbeam}}
\tablewidth{0pt}
\tablehead{
\colhead{Case (comparand)} & \multicolumn2c{\em UHF \em band} & \multicolumn2c{\em L \em band} & \multicolumn2c{\em S \em band}\\
\cline{1-1}
\cline{2-3}
\cline{4-5}
\cline{6-7}
\colhead{} & \colhead{700 MHz} & \colhead{1000 MHz} & \colhead{1000 MHz} & \colhead{1500 MHz} & \colhead{2000 MHz} & \colhead{3000 MHz}
}
\startdata
\colhead{} & \multicolumn6c{Errorbeam (median over all antennas)}\\
\cline{2-7}
Array average & 0.28\% & 0.68\% & 0.22\% & 0.65\% & 0.37\% & 0.55\% \\
Array average shape, per-antenna squint & 0.19\% & 0.18\% & 0.14\% & 0.19\% & 0.29\% & 0.35\% \\
Measurements at 30$^\circ$ elevation & 0.14\% & 0.10\% & 0.09\% & 0.10\% & 0.29\% & 0.37\% \\
Measurements at 0$^\circ$C & 0.10\% & 0.06\% & 0.06\% & 0.06\% & 0.19\% & 0.32\% \\
\hline
\colhead{} & \multicolumn6c{Error in instrumental polarization (median over all antennas)}\\
\cline{2-7}
Array average & 0.37\% & 0.69\% & 0.22\% & 0.89\% & 0.34\% & 0.93\% \\
Array average shape, per-antenna squint & 0.38\% & 0.63\% & 0.20\% & 0.87\% & 0.34\% & 0.92\% \\
Measurements at 30$^\circ$ elevation & 0.21\% & 0.18\% & 0.18\% & 0.21\% & 0.50\% & 0.39\% \\
Measurements at 0$^\circ$C & 0.08\% & 0.10\% & 0.08\% & 0.12\% & 0.35\% & 0.34\% \\
\enddata
\end{deluxetable*}
In Figure \ref{fig:errorbeam_uls} the cross-polarization and subsequently the instrumental polarization of m057 in the \em L \em band is exceptionally large due to the installation of the underperforming l.4046 OMT component. It appears that the polarization performance is more affected by receivers installed than even severe collimation errors. The smaller range of variation in instrumental polarization in UHF than L  (and also S) may be due to the manufacturing tolerance for the OMTs remaining constant and therefore comparatively smaller at lower frequencies.

Figure \ref{fig:instrumental_cross_intensity} illuminates how the instrumental polarization and cross-polarization power is distributed over the mainlobe of the beam. Within the 90\% Stokes I power contour of the beam, better than 1\% instrumental polarization can be expected in the lower half of each band, but this approaches 2\% in the upper third of the \em UHF \em and \em L \em bands. Within the half-power region of the beam up to 3\% instrumental power can be expected in the lower half of the band but this can increase to beyond 10\% in the upper half of the band. While the instrumental polarization performance may be better at some lower frequencies in the \em S \em band than the other bands, the performance is more varied (as it deteriorates more quickly away from the beam center) at several intermittent frequency neighborhoods.

\subsection{Perspective on beam shape variability}
\label{section:perspective_on_beam_shape_variability}
Table \ref{tab:errorbeam} quantifies by how much the nominal (reference) per-antenna beam patterns, at 60$^\circ$ elevation and 15$^\circ$C, change in shape due to a few topical effects. Both Stokes I errorbeam, and the error in instrumental polarization are listed for four cases of interest. These metrics evaluate the maximum power differences over the half-power region of each beam. The median is taken over all antennas (not showing the worst antenna case). One low and one high frequency result are shown per band.

The \em array average \em case refers to the expected error in the beam patterns that occurs when the beams are modeled using a single array average beam for all antennas indiscriminately. Making this simplification in modeling might be computationally advantageous but these results show a marked increase in error in the upper half of each band due to the fact that the waveguide modes are sensitive to manufacturing tolerances and differs significantly per antenna.

In the \em array average shape, per-antenna squint \em case, a single array average beam shape is still assumed, but at each frequency channel, for each antenna, a tailored per-antenna squint offset is used instead of that of the array average. There is a notable improvement in the Stokes I errorbeam in this case compared to using an array average, but the advantage is negligible in terms of the instrumental polarization. Simply customizing the squint pointing in this manner is therefore not particularly effective for improving the modeling of cross-polarization, even though doing so is quite effective in modeling copolarization beams.

The \em measurements at 30\textdegree{} elevation \em case compares the reference beam patterns against the commonly occurring low elevation situation to size up the effect of gravitational loading at commonly occurring low elevations. Likewise, the change in beam shape for \em measurements at 0\textdegree C \em is a bit smaller but remains significant. Environmental effects on the \em S \em band appears to be greater.

\section{Aperture plane results}
\label{section:aperture_plane_results}

For reflector radio telescopes, holography measurements are usually performed in the \em Ku \em (or \em Ka \em) band using satellite beacon signals that are available at limited frequencies and antenna elevations. Aperture plane measurements $A(x,y)$ are obtained by Fourier transform of the complex-valued far-field copolarized patterns $E_{HH}$ and  $E_{VV}$. These aperture plane measurements are used primarily to check that the reflector surfaces of an antenna collectively focus at the feed position, and to diagnose how the feed and individual reflector panels need to be adjusted in order to correct any installation inaccuracies. A complementary technology, commonly photogrammetry, is usually employed beforehand during the assembly phase of an antenna to guide installation adjustments for individual structural components. Because component tolerances compound and since the mechanical and electromagnetic surfaces don't necessarily coincide, holography is performed as an overall system check. Furthermore, subtle gravitational loading differences may occur when more receivers are mounted and the installation may degrade over time, necessitating monitoring which is automated using holography.

Conventionally the \em Ku \em band is preferred over lower observation frequencies: a narrower angular range of scanning is needed to achieve the required resolution. This improves the measurement integrity because antenna structures may physically deform at sufficiently different elevations due to gravitational loading. Furthermore, compared to the much lower observation frequencies used by MeerKAT, the \em Ku \em band suffers much less from frequency-dependent diffraction effects superimposed over the aperture plane in the measurements. However, it is not possible to monitor the antenna structure over the MeerKAT lifetime in the \em Ku \em band because this telescope only supports these feeds in a qualification test setup that is very labour intensive to interchange.

Routine monitoring of collimation and dish surface errors is therefore done in the standard observational bands on celestial targets using the same datasets as for beam shape characterization. Ensuring the viability of transforming the measured beam to the aperture plane, the scan extent is wide enough to include a few sidelobes at the lowest frequencies in the band. The \em L \em band is preferred over the \em S \em band for characterizing environmental effects on collimation due to the availability of higher flux celestial targets that permits shorter duration observation cycles.

\begin{figure}[t]
\centering
\includegraphics[width=\linewidth,trim=1.25cm 1.25cm 1.25cm 0.5cm, clip]{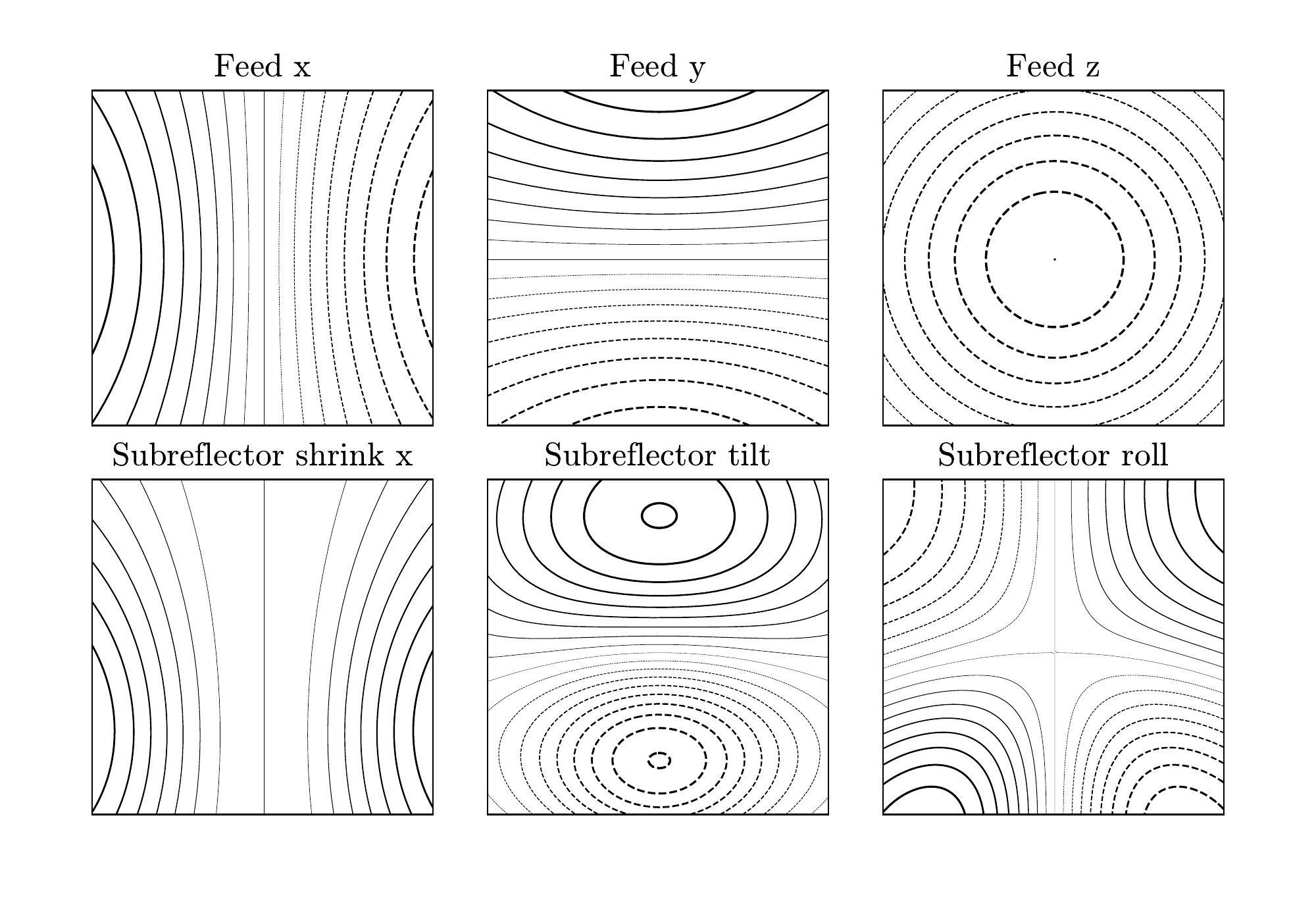}
\caption{Pathlength errors modeled by means of ray tracing are illustrated as contours over the extended aperture (17 m across). The contours are spaced at 0.1 mm per mm (pathlength error per position error) for the first four panels, and 0.1 mm per quarter degree (pathlength error per rotational alignment error) for the last two panels. Negative values are indicated using dashed lines.
\label{fig:pathlengtherrors}}
\end{figure}

\subsection{Effect of collimation errors on pathlength}

At a frequency, $f$, with corresponding wavelength, $\lambda$, the two-dimensionally unwrapped phase angle, $\theta_\mathrm{unwrap}(x,y)$, over the aperture plane $(x,y)$, relates to the pathlength, $p(x,y)$, as:
\begin{equation}
\Delta p(x,y)=\frac{\lambda}{2 \pi} \left(\theta_\mathrm{unwrap}(x,y)- \theta_\mathrm{ref,unwrap}(x,y)\right)
\end{equation}
where $\theta_\mathrm{ref,unwrap}$ includes mainly the phase effects of the feed and can either be estimated through EM simulation or from suitable array average results. If the pathlength errors are small, then the deviation, $\Delta d(x,y)$, normal to the surface of a parabolic reflector can be approximated using the equation
\begin{equation}
\Delta d(x,y)=\frac{\sqrt{(x-x_0)^2+(y-y_0)^2+4 f^2}}{4 f} \Delta p(x,y)
\end{equation}
where $f$ is the focal length of the parabola with its vertex at $(x_0,y_0)$.

\begin{figure*}[t]
\centering
\includegraphics[width=\linewidth,trim=0.5cm 0cm 0cm 0cm, clip]{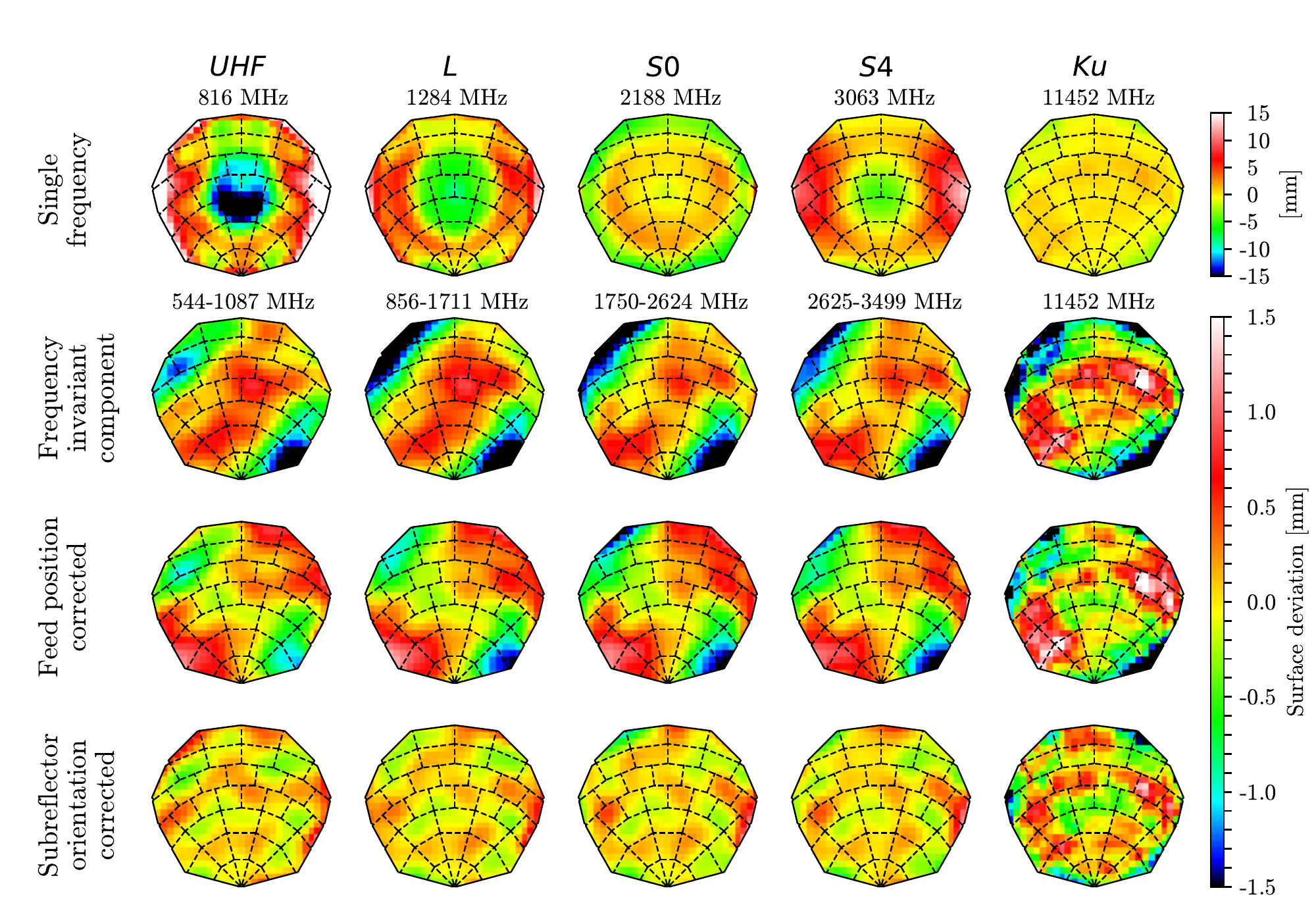}
\caption{Main reflector panel surface errors for antenna m028 are computed at 60\textdegree{} elevation in each of the MeerKAT bands for the H-polarization. The aperture plane phase maps, hence pathlength errors, are expressed as surface deviation errors relative to a parabolic dish surface. The top row shows that frequency dependent features superimposed onto the aperture plane phase are more substantial at lower frequencies and are almost absent in the \em Ku \em band. In the second row, for the \em UHF\em , \em L \em, and \em S \em bands, the frequency invariant phase component over the respective band is shown, revealing phase effects due to collimation errors. In the third row feed positioning phase errors are removed and while in the final row sub-reflector orientation errors are also removed. In the single-frequency \em Ku\em-band result, the small contribution from the feed phase front manifests as a green anticlockwise tilted bowtie pattern in the center of its map, is not removed here. This feature could be removed using either array average information, which is not available in the \em Ku \em band, or less accurately, by using EM simulation results. Note that different feed position errors exist for each band due to their unique mounting errors on the feed indexer. Antenna m028 is chosen in this example because its collimation has remained stable over time and this is the only antenna for which recent \em Ku\em -band measurements are available. The \em Ku \em measurement is dated August 17 2021, at an elevation of 52\textdegree.
\label{fig:compare_surface}}
\end{figure*}

Figure \ref{fig:pathlengtherrors} shows pathlength distortions that occur over the aperture plane as a result of small changes in the geometry of a MeerKAT antenna that commonly occurs. While the equations by \cite{Ruze_1969} could be used to approximate a subset of these effects, the models shown are calculated by means of ray tracing and are used instead to estimate geometric distortions, by fitting these to pathlength measurements. Since the patterns due to feed position offsets are highly correlated to that of sub-reflector offsets (not shown), holography cannot be used to distinguish between such instances. However an adjustment to the feed position can compensate for the effect of a sub-reflector position error.

Simple linear phase gradients (not shown) from left to right or top to bottom across the aperture commonly occur due to pointing errors. These are removed during the fit alongside any phase offsets.

It is interesting that the sub-reflector roll (a small axial rotation of the sub-reflector as viewed from the feed) results in a saddle-shaped pathlength error pattern across the aperture. For MeerKAT, the final installation adjustments of the sub-reflectors proved to be difficult during photogrammetry shoots due to its flexibility and limited favorable wind conditions. Despite the mounting of additional support struts for increased rigidity it remains ambiguous to distinguish between a sub-reflector roll or an actual shape error for some antennas.

Although small geometric errors can be estimated reliably by fitting these models to pathlength measurements, amplitude and polarization effects on the aperture are less easy to predict. It is therefore not accurate to reverse predict hypothetical effects of geometric distortions back onto the beam shape directly from these pathlength models using the inverse Fourier transform. For such predictions, electromagnetic analysis techniques are advised.

\subsection{Aperture plane results for the different bands}

The first row in Figure \ref{fig:compare_surface} shows the measured phase maps over the aperture of antenna m028 at five different frequencies for H-polarization feeds near the center of each band. Despite comparable environmental conditions in each case, these results are vastly different due to superimposed frequency-dependent feed phase pattern effects that dominate over the collimation effects at the lower frequency bands. Phase maps that change rapidly as a function of frequency are predominantly a consequence of the electromagnetic characteristics of the feed, the OMT (in the upper half of each band) and constructive/destructive interference of multi-path transmission between the feed horn and reflectors, rather than reflector collimation errors. 

By subtracting the feed phase pattern and averaging out the frequency dependency over a wide bandwidth, as shown in the second row of the figure over each respective band, a much more stable result is revealed. In the case of the \em Ku \em band no such averaging is done because a single frequency satellite beacon was used and the frequency effects are much smaller. 

Each band's feed has somewhat independent feed position errors. When these errors are solved for and removed, as shown in the third row, the results become more similar for all the bands. However in the \em Ku \em band a tilted bowtie shaped feature starts to appear that is due to the feed's polarization response to the beacon signal. The polarization response is removed for the \em UHF\em, \em L \em and \em S \em bands where the array average response is available and can summarily be subtracted. 

In the final row of the figure, the sub-reflector orientation pathlength effect is also removed, revealing an estimate of main reflector panel errors. It is clear in this example that the feed position and sub-reflector orientation errors distort flatness of the aperture phase more than main reflector panel errors. A sought after flat phase (i.e. zero pathlength error) over the aperture results in the maximum phase efficiency. 

\subsection{Tracking collimation errors over time}

\begin{figure}[t]
\centering
\includegraphics[width=\linewidth,trim=1.1cm 0.25cm 2cm 1.25cm, clip]{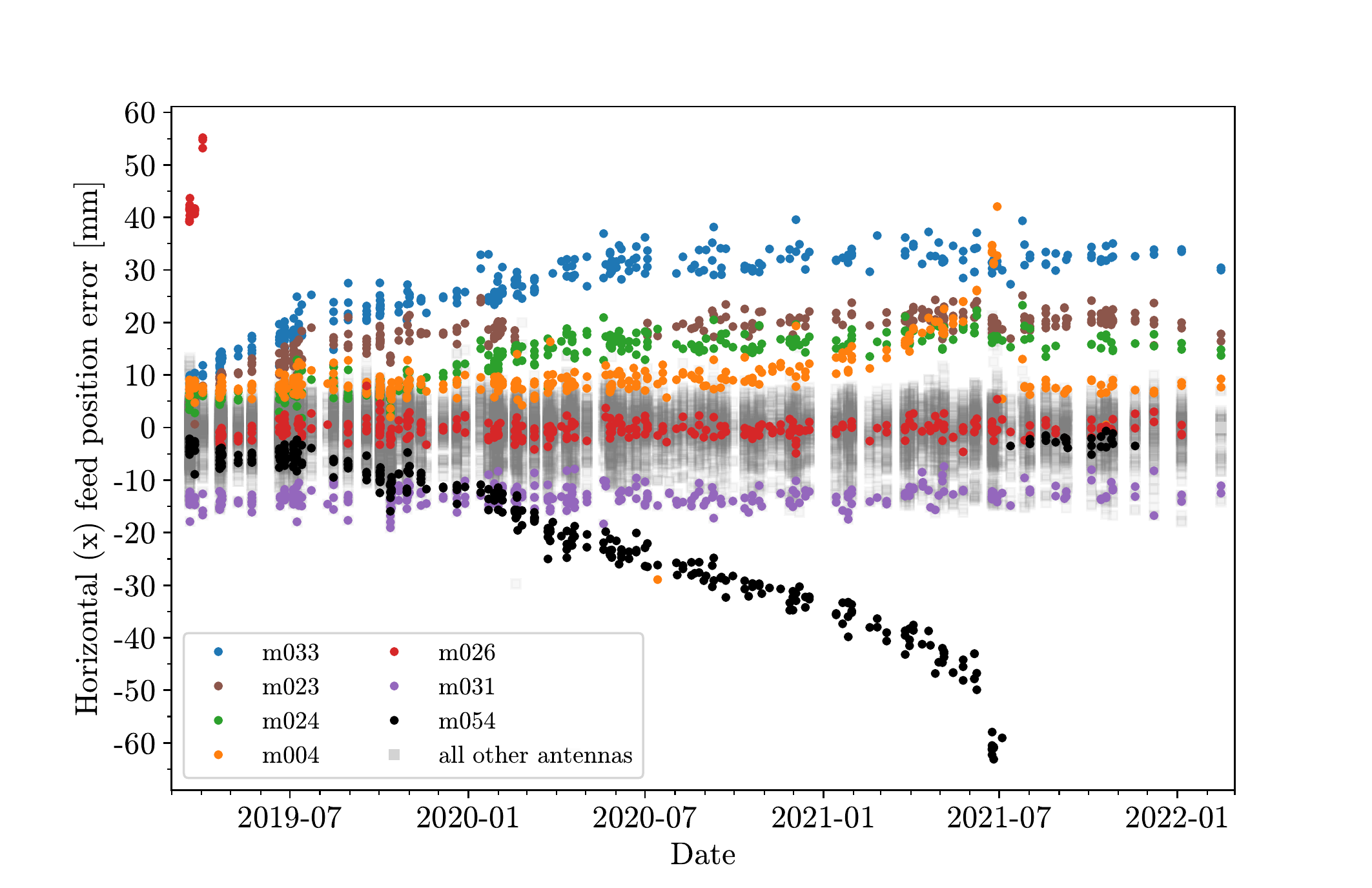}
\caption{Collimation errors due to indexer misalignment and slippage. Although the extreme slippage of m054 and m004 was corrected  in July 2021, m033 remains in significant error. Note that antenna m026 suffered a faulty receiver l.4058 installation that was swiftly removed. After servicing, this receiver was installed onto antenna m005 and thereafter m000 at later dates without recurring problems.
\label{fig:collimation_date}}
\end{figure}

The measurement of collimation errors over the installation's history in Figure \ref{fig:collimation_date} shows large and deteriorating feed position errors for a few antennas. While some antennas such as m031 appear to have had a significant yet stable installation error, m004 and m054 deteriorated drastically until it was fixed in July 2021. In these cases, the encoder couplings of the affected feed indexers creeped in their mounted positions due to repetitive mechanical strains and an insufficiently tight fit. This resulted in a physical discrepancies between actual and commanded positions. Some other antennas like m033 has not yet been corrected resulting in ongoing poor errorbeam performance illustrated in Section \ref{section:errorbeam}.

\begin{figure*}[t]
\centering
\includegraphics[width=\linewidth,trim=4cm 0cm 3.35cm 0cm, clip]{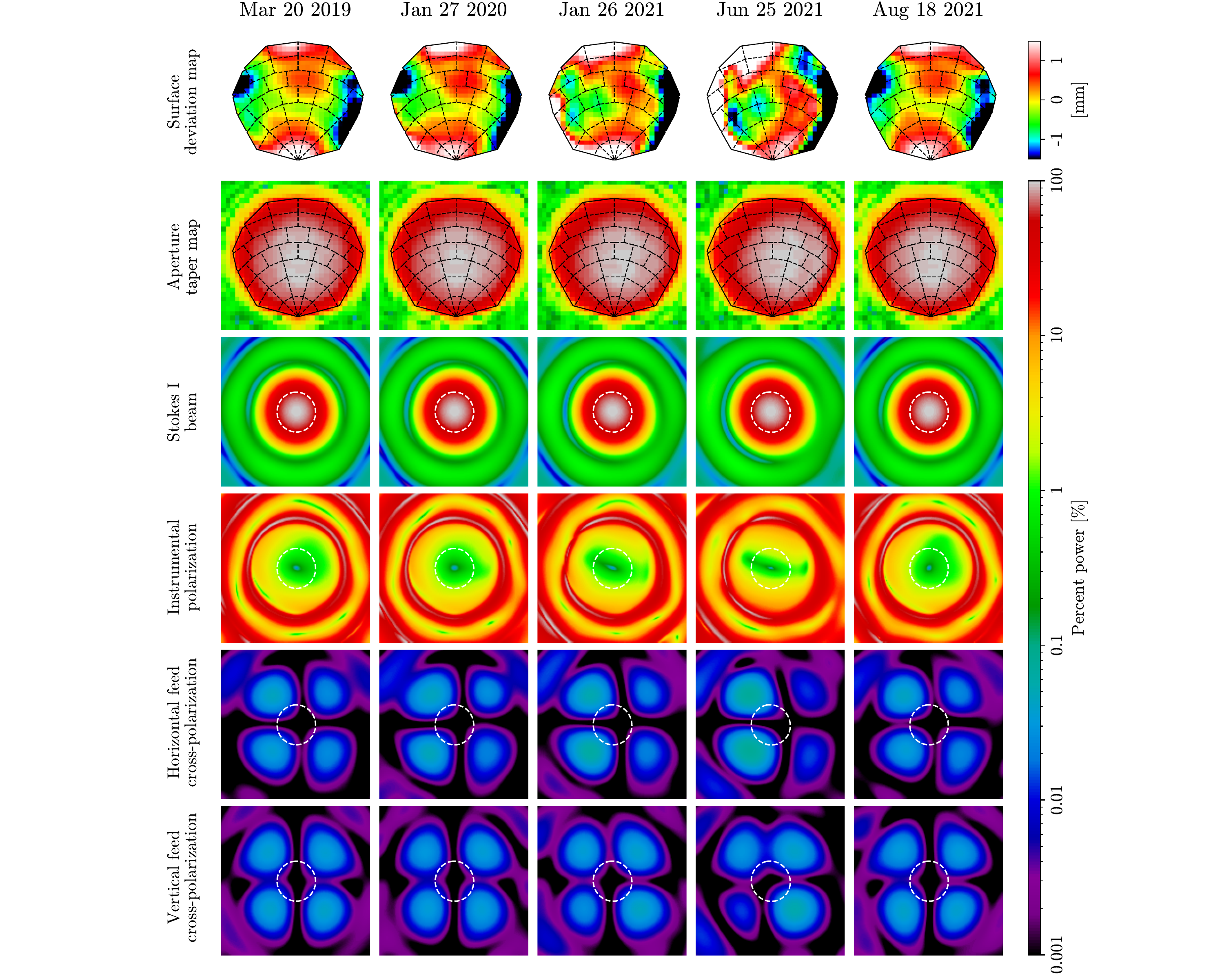}
\caption{Effect of feed indexer table rotation error due to encoder mount slippage on the beams of antenna m054 with receiver l.4050 installed during the entire timeframe. A 25$^\circ$ rotation of the fitted ellipse at the half-power point (dashed white lines) is measured on the Stokes I beam of 25 June 2021. The fault was corrected during July 2021.
\label{fig:beams_date}}
\end{figure*}

Figure \ref{fig:beams_date} shows in a progression of time how severely various beam shape and aperture plane measurements of m054 were affected by the feed indexer creep effect. The first row shows how the pathlength error is affected. The second row shows that the power distribution over the aperture becomes increasingly accentuated towards the right-hand side up to June 2021. The third and fourth rows show that the Stokes I beam pattern and instrumental polarization become elongated diagonally, while the last two rows show that the cross-polarization beam response is also characteristically affected at lower levels of power.

\subsection{Elevation and temperature effects on collimation}

In addition to collimation errors that are unique to antennas individually, there are also collimation error trends which are shared amongst antennas due to environmental factors. The elevation effect on collimation is the most impactful as gravitational loading changes the shape of the reflectors. Skipping the more complicated (multi-dimensional) daytime sun-angle effects which are avoidable and not quantified here, the next most noticeable effect is that of the ambient temperature. Although wind loading plays a further role, that effect seems to be smaller and there is too little data given the reliability of the wind measurement sensors and dimensionality (wind loading has a dependence on wind speed, direction and antenna elevation) to report conclusively on the wind effect at this stage.

\begin{figure}[t]
\centering
\includegraphics[width=\linewidth,trim=0.25cm 0cm 0.25cm 0cm, clip]{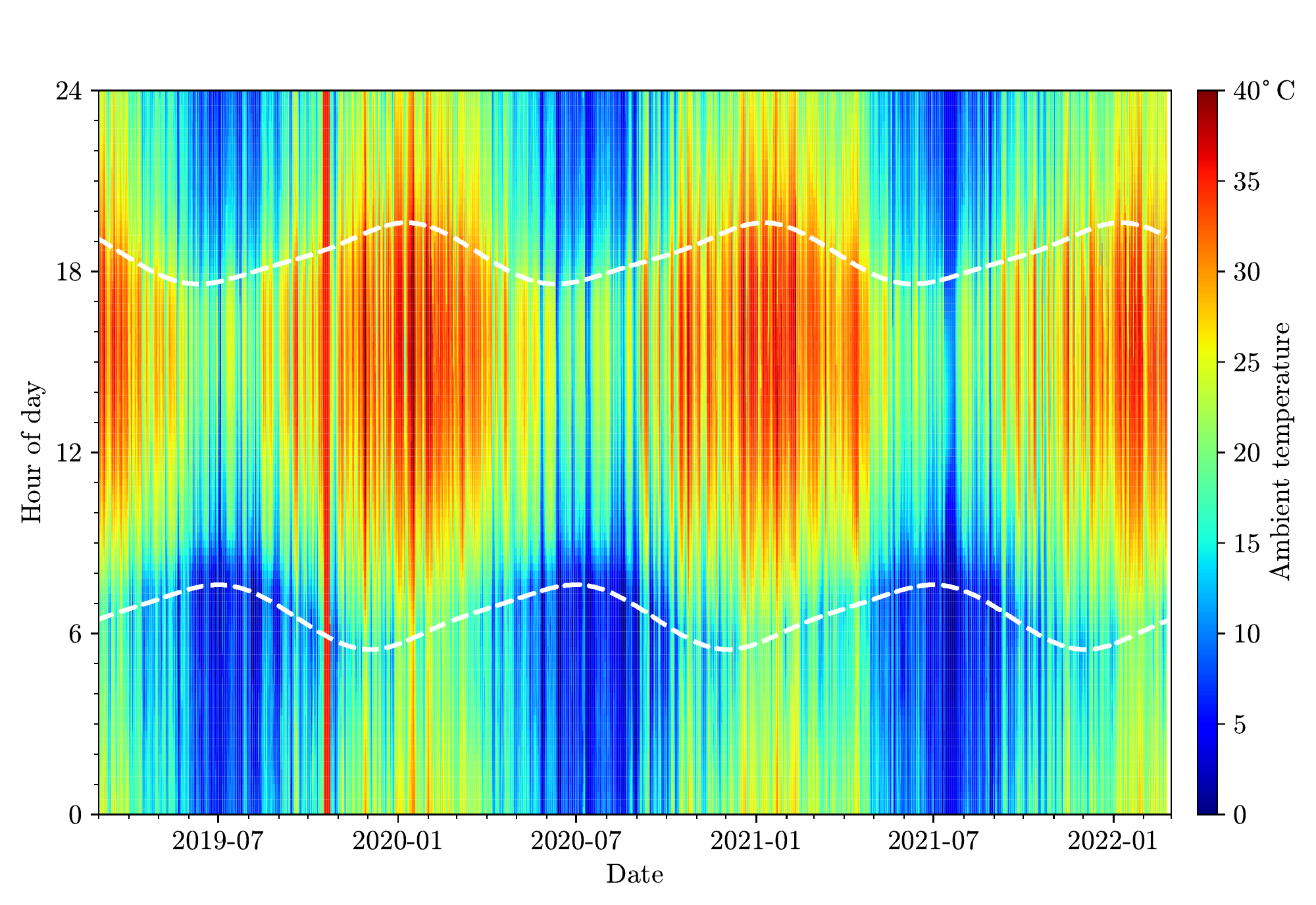}
\caption{Daily ambient temperature variability experienced at the MeerKAT site in the Karoo over the years from 2019 to 2022. An overnight cooldown of air temperature by 10$^\circ$C is most common. Sunrise and sunset times are indicated using white dashed lines. A few cases of sensor failure can be seen, e.g. 16--23 October 2019 and 22--23 July 2021.
\label{fig:temperature}}
\end{figure}

Figure \ref{fig:temperature} shows that the ambient temperatures at the MeerKAT site varies between 0$^\circ$C and 40$^\circ$C depending on the season, and it is common for the temperature to drop by about 10$^\circ$C during an observation that runs overnight at any time of the year.

\begin{figure*}[t]
\centering
\includegraphics[width=\linewidth,trim=0.8cm 0.25cm 0.5cm 0.5cm, clip]{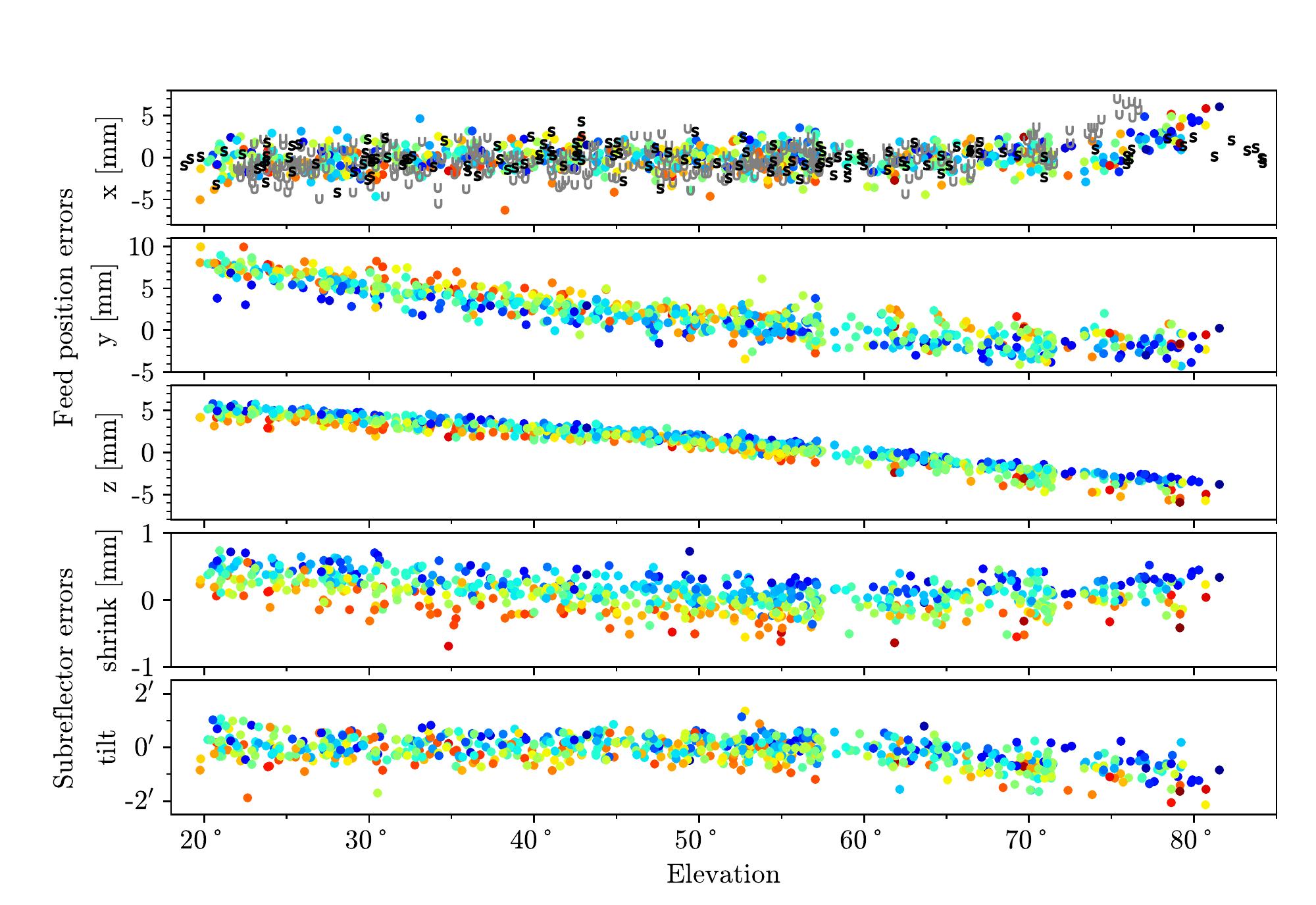}
\caption{The impact of environmental conditions, notably elevation and temperature, on the measured collimation errors in the \em L \em band. In the top panel only, results for the \em UHF \em and \em S \em bands (marked using \texttt{U} and \texttt{S} respectively) are included to show behavioral differences at high elevations amongst the bands in the horizontal direction due to asymmetrical gravitational loading of the indexer table. The applicable temperature color scale is shared in Figure \ref{fig:temperature}.
\label{fig:collimation_elevation}}
\end{figure*}

The elevation and temperature effects on collimation are illustrated in Figure \ref{fig:collimation_elevation}. Most striking is a bulging of the main reflector due to gravitational loading that results in the feed moving out of focus by up to 10 mm in the vertical direction at low elevation angles, and by 5 mm towards or away from the sub-reflector. This is in good agreement with the original structural design analysis.

Curiously, the feed indexer and sub-reflector support structure appears to lean to the side at high elevations depending on which feed is indexed. This is due to an asymmetric weight loading on the feed indexer that strains the feed indexer and sub-reflector support structures at high elevations depending on its center of gravity. Above 75\textdegree{} elevation the indexer is most balanced across the symmetry plane of reflector optics in the \em S \em band case, and least balanced when indexed for the \em UHF \em band. Fewer science observations require elevations above 75$^\circ$ since high elevations service a small patch of the sky.

\begin{figure*}[t]
\centering
\includegraphics[width=\linewidth,trim=0.12cm 0cm 0.12cm 1.5cm, clip]{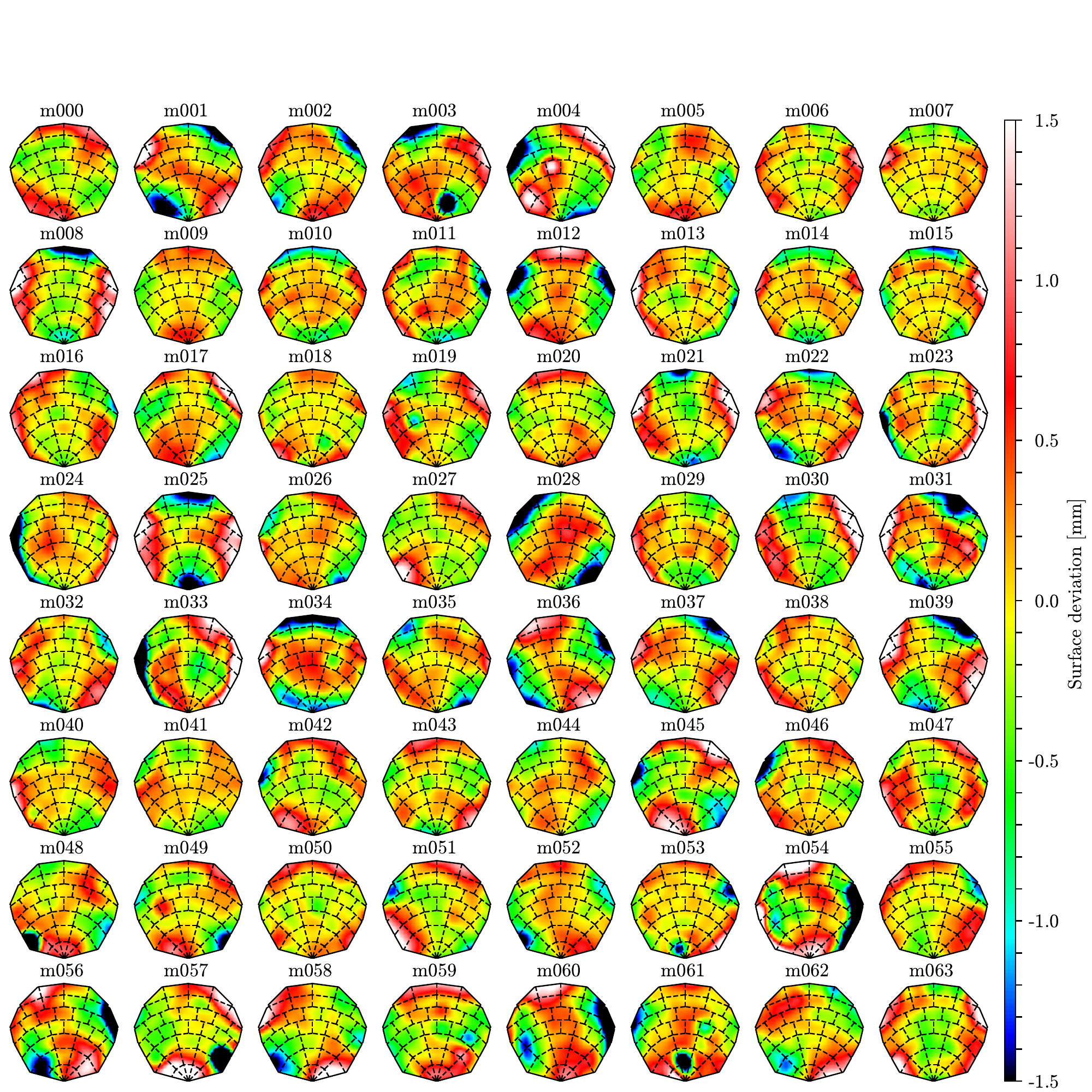}
\caption{Surface and collimation errors for all MeerKAT antennas measured at 60$^\circ$ elevation using \em L\em-band feeds as averaged over the time period from January 2020 to January 2022. Contributions due to feed positioning and sub-reflector orientation as well as main reflector panel errors are included in these results, whereas antenna pointing error effects are removed. These are as for m028 in the second row of Figure \ref{fig:compare_surface}.
\label{fig:surface_l}}
\end{figure*}

\subsection{Pathlength error maps for all antennas}

Pathlength error maps measured using the \em L\em-band feeds are shown for all MeerKAT antennas in Figure \ref{fig:surface_l}. These results show the collimation effects due to any reflector misalignment including main reflector panel, sub-reflector orientation and feed position errors. The display of this combination is preferred because it shows how the different effects size up against each other. 

The dark blue colored panels for m003 and m061 (near the bottom center in both cases) show sizable main reflector panel errors in the direction away from the sub-reflector. The white colored panel of m004 (near the center) shows such an error towards the sub-reflector. Prior to 11 November 2019 when a loose stud was fastened (stud 442 unrestricted by 8.75 mm, not shown), a corresponding panel error appeared to move depending on the elevation in the holography results of m050.

For antennas, such as m023, m024, m033, m054 where black/blue and red/white streaks persist towards the left and right hand edges respectively (and vice versa), there are significant rotational feed indexer errors. A saddle-shaped pattern such as for m039 can be explained by an $\sim -0.5^\circ$ sub-reflector roll.

Holography analysis is traditionally used to make corrections to both feed position errors and surface errors, however this is infrequently done for MeerKAT. Instead, the `as-is' situation is routinely monitored for engineering purposes. Individual panel errors have a small effect on performance compared to unavoidable factors such as gravitational loading due to elevation.

Finally, Table \ref{tab:collimation_errors} reports on the status of measured collimation errors for all antennas, as averaged over the 2020 to 2022 period. Feed indexer offsets are systematically being resolved during maintenance cycles.

\section{Conclusion}

This paper presented the measured primary beam characteristics for the MeerKAT \em UHF\em, \em L \em and \em S \em wide-band receivers. In agreement with earlier work, it is found that the varied activation of higher-order waveguide modes, rather than commonly occurring errors in the reflector geometry, is responsible for the dominant perturbations to the primary beam shape in the upper half of each band. Furthermore, additional data suggest that the associated frequency and polarization dependent pointing (squint) and ellipticity are pervasive phenomena amongst \em different \em OMT designs and are likely to affect wide bandwidth waveguide feeds more generally. This is of wider interest than only to MeerKAT science data users because the \em L\em-band OMT design is adapted for use in the upcoming ngVLA, as well SKA telescopes. As such, very similar features are expected to appear in their affected bands. 

These effects occur above the theoretical cut-off frequency associated with the TE21 and TM11 modes and their presence can be anticipated for given waveguide diameters. However, the exact manifestation of these characteristics are sensitive to designs and manufacturing tolerances, and can dominate the departures of the primary beam shape from that of the array average while straining the instrumental polarization performance. It is impractical to predict accurately these differences on a per-receiver basis by means of EM simulations.

Success in the estimation of collimation errors were demonstrated at low observational frequencies that are overwhelmed by diffraction and feed phase effects. The same measurements are used to monitor the condition of the antennas as to characterize its primary beams.

The \em UHF \em and \em L \em bands, which have been studied for a longer period of time than the \em S \em band, show significant changes in beam shapes developing due to feed indexer slippage for some antennas. Likewise, the beam shapes are impacted identifiably when receiver units are exchanged amongst antennas in the array.

The accompanying data release shares a time-averaged snapshot of the full Jones primary beam results for all MeerKAT antennas and bands. Apart from post-imaging primary beam corrections, and imaging simulation studies, these per-antenna primary beam measurements may be useful in the application of A-projection imaging techniques \citep{Bhatnagar_2013}, and high dynamic range imaging efforts \citep{Mitra_2015}. Ultimately the fidelity of off-boresight interferometric polarization observations of the telescope stands to be improved. Future work will address primary beam modeling in the face of configuration changes (which receiver is installed), time (collimation decay), and environmental conditions (elevation and temperature).

\begin{acknowledgments}
This paper was reviewed internally by Fernando Camilo, Ludwig Schwardt and Adriaan Peens-Hough. Thanks to Sarah Buchner and Tumelo Mangena for facilitating the scheduling and capture of numerous observations, and to SARAO staff for keeping the telescope operational.

The MeerKAT telescope is operated by the South African Radio Astronomy Observatory, which is a facility of the National Research Foundation, an agency of the Department of Science and Innovation. This work has made use of the ``MPIfR \em S\em -band receiver system" designed, constructed and maintained by funding of the MPI f\"{u}r Radioastronomy and the Max-Planck-Society.
\end{acknowledgments}

\begin{deluxetable*}{@{\extracolsep{4pt}}ccccccccccccccccc}[!t]
\tabletypesize{\scriptsize}
\renewcommand{\arraystretch}{0.8}
\tablenum{4}
\tablecaption{A summary of measured collimation errors at 60\textdegree{} elevation and 15\textdegree C. In the case of the \em UHF \em and \em L \em bands, the average results are listed over the time period from January 2020 until January 2022. For the \em S \em band, the average is taken over the range from February 2021 until February 2022. The (parallactic) dipole orientations, Stokes I beam ellipticity orientation, and sub-reflector roll error are listed under $\theta$, $\phi$ and $\Psi$ respectively.}
\label{tab:collimation_errors}
\tablewidth{0pt}
\tablehead{
\colhead{Antenna} & \multicolumn5c{\em UHF \em band feed} & \multicolumn5c{\em L \em band feed} & \multicolumn5c{\em S \em band feed} & \colhead{Sub-reflector}\\
\cline{1-1}
\cline{2-6}
\cline{7-11}
\cline{12-16}
\cline{17-17}
\colhead{} & \colhead{x} & \colhead{y} & \colhead{z} & \colhead{$\theta$} & \colhead{$\phi$} & \colhead{x} & \colhead{y} & \colhead{z} & \colhead{$\theta$} & \colhead{$\phi$} & \colhead{x} & \colhead{y} & \colhead{z} & \colhead{$\theta$} & \colhead{$\phi$} & \colhead{$\Psi$}
}
\startdata
m000 & -3 & -1 & 1 & 0.8\textdegree & 0\textdegree & -3 & -2 & 2 & 0.2\textdegree & -1\textdegree & -1 & -3 & 2 & 1.8\textdegree & -2\textdegree & 0.3\textdegree\\
m001 & -3 & 1 & 0 & 0.4\textdegree & -2\textdegree & -1 & 1 & -1 & 0.6\textdegree & -3\textdegree & -1 & -2 & -1 & -0.6\textdegree & -3\textdegree & -0.7\textdegree\\
m002 & -9 & 4 & 0 & 0.2\textdegree & -2\textdegree & -7 & 4 & 0 & 0.7\textdegree & -1\textdegree & -7 & 2 & 1 & -0.7\textdegree & -7\textdegree & -0.4\textdegree\\
m003 & 3 & 7 & -1 & 0.1\textdegree & 3\textdegree & 6 & 5 & -1 & 0.9\textdegree & 2\textdegree & 0 & 2 & -2 & -0.8\textdegree & 6\textdegree & 0.5\textdegree\\
m004 & 12 & -4 & -1 & -1.0\textdegree & 1\textdegree & 13 & -5 & -1 & -0.7\textdegree & 2\textdegree & 11 & -4 & -2 & -1.6\textdegree & 5\textdegree & 0.6\textdegree\\
m005 & -1 & 4 & 1 & -0.6\textdegree & -1\textdegree & -1 & 3 & 0 & -1.3\textdegree & -1\textdegree & 2 & 3 & 3 & -0.6\textdegree & 0\textdegree & 0.1\textdegree\\
m006 & -3 & 2 & 1 & -0.1\textdegree & 0\textdegree & 1 & 1 & 1 & -3.2\textdegree & 0\textdegree & -1 & -2 & 2 & 0.8\textdegree & -1\textdegree & 0.0\textdegree\\
m007 & 3 & 2 & 0 & 1.3\textdegree & 0\textdegree & -1 & 2 & 0 & 0.4\textdegree & 0\textdegree & -2 & 4 & -25 & -0.4\textdegree & -2\textdegree & -0.1\textdegree\\
m008 & -5 & 5 & 2 & 0.5\textdegree & 0\textdegree & -1 & 4 & 2 & -0.3\textdegree & 0\textdegree & -23 & 24 & -69 & -2.7\textdegree & 0\textdegree & -0.1\textdegree\\
m009 & 0 & -2 & -1 & 1.2\textdegree & -1\textdegree & 1 & -2 & 0 & 0.0\textdegree & 0\textdegree & 2 & 0 & 0 & -0.0\textdegree & -1\textdegree & 0.0\textdegree\\
m010 & -1 & 2 & -2 & -0.3\textdegree & 0\textdegree & 1 & 1 & -2 & 0.7\textdegree & -1\textdegree & -1 & 0 & -1 & 0.5\textdegree & 0\textdegree & 0.0\textdegree\\
m011 & 9 & -7 & -3 & -1.6\textdegree & 5\textdegree & -11 & -8 & -1 & 0.7\textdegree & -3\textdegree & 6 & -5 & -1 & -0.2\textdegree & 10\textdegree & 0.3\textdegree\\
m012 & -1 & 3 & -1 & -0.6\textdegree & 1\textdegree & -1 & 0 & -1 & 1.0\textdegree & 0\textdegree & -1 & 2 & -1 & 0.3\textdegree & 2\textdegree & 0.2\textdegree\\
m013 & -5 & -1 & 0 & 0.4\textdegree & -4\textdegree & -9 & 0 & 1 & -0.3\textdegree & -5\textdegree & -10 & 2 & 1 & -1.0\textdegree & -10\textdegree & -0.0\textdegree\\
m014 & -3 & -1 & -2 & 0.5\textdegree & 0\textdegree & -1 & 3 & -1 & -0.6\textdegree & 0\textdegree & -3 & 3 & -2 & -0.2\textdegree & -1\textdegree & -0.1\textdegree\\
m015 & 4 & 2 & 0 & 0.6\textdegree & 1\textdegree & 3 & 3 & 0 & 1.0\textdegree & 3\textdegree & 0 & 1 & 0 & -2.0\textdegree & 2\textdegree & 0.0\textdegree\\
m016 & -7 & 0 & 1 & -0.7\textdegree & -3\textdegree & -6 & 1 & 1 & -0.6\textdegree & -3\textdegree & -4 & 1 & 0 & 0.8\textdegree & -6\textdegree & -0.2\textdegree\\
m017 & 1 & -9 & 1 & -0.7\textdegree & 1\textdegree & 2 & -11 & 0 & -1.1\textdegree & 1\textdegree & 1 & -10 & 1 & 0.1\textdegree & 3\textdegree & 0.3\textdegree\\
m018 & 3 & 4 & 0 & -1.0\textdegree & 0\textdegree & 2 & 2 & 0 & -0.2\textdegree & 0\textdegree & 1 & 0 & 0 & 0.1\textdegree & 1\textdegree & 0.1\textdegree\\
m019 & -2 & 0 & 2 & 0.0\textdegree & -1\textdegree & -2 & -1 & 1 & 0.4\textdegree & 0\textdegree & -4 & 0 & 1 & 0.5\textdegree & 0\textdegree & 0.3\textdegree\\
m020 & -1 & 1 & 1 & 0.6\textdegree & 0\textdegree & -2 & -3 & 0 & -0.4\textdegree & 0\textdegree & -4 & -2 & 0 & 0.3\textdegree & -1\textdegree & -0.1\textdegree\\
m021 & 1 & -6 & 0 & -0.6\textdegree & 0\textdegree & 1 & -6 & 2 & 0.2\textdegree & -1\textdegree & -1 & -8 & 2 & -2.8\textdegree & 0\textdegree & 0.2\textdegree\\
m022 & 2 & 4 & 2 & -0.0\textdegree & 1\textdegree & 5 & 1 & -1 & -1.2\textdegree & 1\textdegree & 1 & 1 & 0 & 2.2\textdegree & 0\textdegree & -0.4\textdegree\\
m023 & 14 & 2 & 0 & 1.5\textdegree & 7\textdegree & 18 & 2 & 2 & 1.3\textdegree & 8\textdegree & 14 & 1 & 2 & 0.4\textdegree & 14\textdegree & -0.1\textdegree\\
m024 & 14 & -1 & 0 & 0.1\textdegree & 7\textdegree & 16 & -3 & -2 & -1.0\textdegree & 8\textdegree & 11 & -2 & 0 & -1.3\textdegree & 13\textdegree & -0.0\textdegree\\
m025 & -3 & 8 & 0 & 1.4\textdegree & 0\textdegree & -4 & 7 & 1 & -0.1\textdegree & 0\textdegree & -5 & 6 & 1 & 0.2\textdegree & -5\textdegree & 0.0\textdegree\\
m026 & -2 & -1 & 0 & 0.2\textdegree & -1\textdegree & -2 & -1 & -1 & 1.0\textdegree & -1\textdegree & -1 & 1 & 0 & 0.2\textdegree & 0\textdegree & 0.3\textdegree\\
m027 & -3 & -1 & 0 & 1.6\textdegree & 0\textdegree & 1 & -1 & 1 & -0.1\textdegree & 0\textdegree & 0 & -2 & 1 & 2.8\textdegree & 0\textdegree & 0.4\textdegree\\
m028 & -2 & -1 & -3 & -0.2\textdegree & -1\textdegree & -1 & 3 & -4 & 0.4\textdegree & 0\textdegree & 0 & 3 & -3 & -1.9\textdegree & 2\textdegree & 0.5\textdegree\\
m029 & -3 & 2 & -1 & 0.1\textdegree & -1\textdegree & -5 & 2 & 0 & -0.6\textdegree & -3\textdegree & -5 & 2 & -1 & 0.1\textdegree & -5\textdegree & -0.0\textdegree\\
m030 & -1 & -6 & 2 & -0.8\textdegree & -1\textdegree & 6 & -2 & 3 & 0.3\textdegree & 2\textdegree & 5 & -5 & 3 & -2.6\textdegree & 8\textdegree & 0.4\textdegree\\
m031 & -12 & 0 & -2 & -0.8\textdegree & -3\textdegree & -12 & 1 & -1 & 0.6\textdegree & -3\textdegree & -10 & -2 & -2 & 0.3\textdegree & -2\textdegree & -0.5\textdegree\\
m032 & -1 & 1 & 1 & 1.2\textdegree & 0\textdegree & -2 & 2 & 1 & 1.1\textdegree & 0\textdegree & -4 & 1 & 3 & 0.4\textdegree & -2\textdegree & -0.3\textdegree\\
m033 & 24 & 0 & 2 & 2.1\textdegree & 12\textdegree & 30 & -1 & 2 & -0.2\textdegree & 13\textdegree & 24 & 1 & 2 & -2.3\textdegree & 22\textdegree & 0.2\textdegree\\
m034 & -4 & 0 & -1 & -0.7\textdegree & 1\textdegree & 2 & 1 & -3 & -0.7\textdegree & 1\textdegree & 1 & -1 & -3 & -0.5\textdegree & -2\textdegree & -0.2\textdegree\\
m035 & 2 & -2 & -1 & 0.0\textdegree & -1\textdegree & -1 & 1 & -1 & 0.8\textdegree & -1\textdegree & 1 & -1 & 0 & -1.0\textdegree & 3\textdegree & 0.4\textdegree\\
m036 & 5 & -7 & -2 & 0.3\textdegree & 3\textdegree & 3 & -6 & -1 & 0.8\textdegree & 2\textdegree & 1 & -8 & -1 & -0.9\textdegree & -2\textdegree & -0.6\textdegree\\
m037 & 0 & 3 & 0 & 0.7\textdegree & 0\textdegree & 0 & 6 & 0 & -1.0\textdegree & 1\textdegree & 2 & 4 & 0 & -0.5\textdegree & -1\textdegree & -0.4\textdegree\\
m038 & -2 & 3 & -1 & -1.7\textdegree & -2\textdegree & -3 & 4 & 1 & 0.7\textdegree & -3\textdegree & 0 & 4 & 1 & -2.7\textdegree & -3\textdegree & 0.0\textdegree\\
m039 & -1 & 0 & 0 & -1.0\textdegree & 1\textdegree & -6 & 1 & 1 & 0.8\textdegree & 1\textdegree & -5 & 0 & 1 & -1.5\textdegree & -1\textdegree & -0.5\textdegree\\
m040 & 2 & 2 & 1 & 0.1\textdegree & 0\textdegree & 0 & 2 & 0 & -0.4\textdegree & 0\textdegree & -2 & 1 & 0 & 0.8\textdegree & -4\textdegree & 0.2\textdegree\\
m041 & 0 & -4 & -1 & 0.5\textdegree & 0\textdegree & 1 & -3 & 0 & -0.5\textdegree & 0\textdegree & -1 & -1 & 0 & 0.9\textdegree & -3\textdegree & 0.1\textdegree\\
m042 & 3 & -2 & 1 & 0.4\textdegree & 0\textdegree & 1 & 0 & 2 & -0.9\textdegree & 0\textdegree & -2 & -1 & 2 & -1.3\textdegree & 2\textdegree & 0.2\textdegree\\
m043 & 1 & -6 & -1 & 0.9\textdegree & 0\textdegree & 2 & -4 & 1 & -0.5\textdegree & 0\textdegree & 0 & -4 & 0 & -0.4\textdegree & 0\textdegree & -0.2\textdegree\\
m044 & -2 & 0 & 0 & 0.5\textdegree & -1\textdegree & -2 & 1 & -1 & 0.4\textdegree & 0\textdegree & -1 & 0 & 0 & 0.2\textdegree & -2\textdegree & 0.3\textdegree\\
m045 & 1 & -5 & 0 & -1.2\textdegree & -1\textdegree & 2 & -5 & 1 & -0.5\textdegree & 0\textdegree & 1 & -2 & 1 & -1.4\textdegree & 0\textdegree & 0.5\textdegree\\
m046 & 3 & 0 & -1 & 0.1\textdegree & 0\textdegree & 3 & 2 & 0 & -0.7\textdegree & 1\textdegree & 2 & 4 & -1 & -0.4\textdegree & 3\textdegree & 0.2\textdegree\\
m047 & 2 & -4 & 0 & -0.0\textdegree & 4\textdegree & 5 & -1 & 2 & -0.7\textdegree & 2\textdegree & 7 & 0 & 1 & 0.4\textdegree & 3\textdegree & 0.1\textdegree\\
m048 & 2 & -3 & 0 & 1.4\textdegree & 0\textdegree & 3 & -3 & -1 & 0.5\textdegree & 0\textdegree & 2 & -3 & 1 & 0.6\textdegree & -3\textdegree & 0.1\textdegree\\
m049 & 2 & -1 & 2 & -1.2\textdegree & -1\textdegree & 2 & -3 & 0 & -0.3\textdegree & 0\textdegree & 3 & -1 & 1 & 1.6\textdegree & -2\textdegree & 0.3\textdegree\\
m050 & 1 & -6 & 0 & -0.5\textdegree & 0\textdegree & 1 & -4 & 2 & 0.4\textdegree & 0\textdegree & 2 & -1 & 1 & 1.6\textdegree & -2\textdegree & -0.1\textdegree\\
m051 & -7 & -5 & 1 & 0.1\textdegree & 0\textdegree & -3 & -3 & 1 & -0.2\textdegree & -1\textdegree & -5 & -2 & 1 & -1.7\textdegree & 1\textdegree & 0.3\textdegree\\
m052 & 1 & -5 & -2 & -0.6\textdegree & 1\textdegree & 4 & -2 & -1 & -0.1\textdegree & 1\textdegree & 6 & 5 & 3 & -0.5\textdegree & 1\textdegree & -0.3\textdegree\\
m053 & 4 & 1 & -1 & 0.0\textdegree & -1\textdegree & 1 & 1 & 0 & -0.8\textdegree & -2\textdegree & 0 & 1 & -2 & -0.7\textdegree & -7\textdegree & -0.2\textdegree\\
m054 & -17 & 3 & 0 & 1.6\textdegree & -7\textdegree & -27 & 0 & -1 & 0.2\textdegree & -12\textdegree & -6 & 4 & -2 & 4.8\textdegree & -9\textdegree & -0.0\textdegree\\
m055 & -5 & 0 & 3 & -1.6\textdegree & 1\textdegree & -6 & 0 & 1 & 0.5\textdegree & 1\textdegree & -6 & 2 & 2 & -0.3\textdegree & 0\textdegree & -0.3\textdegree\\
m056 & -1 & -3 & 1 & -0.1\textdegree & -1\textdegree & -6 & -5 & -1 & -0.3\textdegree & -2\textdegree & 1 & -3 & 1 & 0.4\textdegree & -3\textdegree & -0.7\textdegree\\
m057 & -2 & -4 & 3 & 0.9\textdegree & 0\textdegree & -1 & -6 & 1 & 0.1\textdegree & -1\textdegree & 4 & -1 & 3 & 1.6\textdegree & -2\textdegree & 0.4\textdegree\\
m058 & -3 & 2 & 1 & -0.3\textdegree & 0\textdegree & -3 & 1 & 1 & -0.5\textdegree & 0\textdegree & -4 & 1 & 1 & 0.2\textdegree & -3\textdegree & -0.6\textdegree\\
m059 & 2 & -4 & 1 & -1.5\textdegree & 0\textdegree & 3 & -3 & 1 & 0.4\textdegree & 1\textdegree & 2 & -2 & 3 & -1.0\textdegree & 6\textdegree & -0.2\textdegree\\
m060 & -12 & 0 & -1 & 0.9\textdegree & -7\textdegree & -10 & -2 & -2 & -0.4\textdegree & -4\textdegree & -10 & 1 & -1 & -2.1\textdegree & -7\textdegree & -0.5\textdegree\\
m061 & 4 & 9 & 0 & 0.8\textdegree & 0\textdegree & 5 & 8 & -1 & 0.4\textdegree & 2\textdegree & 6 & 6 & 0 & 0.6\textdegree & 4\textdegree & 0.0\textdegree\\
m062 & -2 & 3 & 1 & -1.7\textdegree & 1\textdegree & 1 & 6 & 0 & 0.9\textdegree & 1\textdegree & -1 & 6 & 0 & -0.4\textdegree & 2\textdegree & -0.5\textdegree\\
m063 & -1 & 1 & 3 & -0.8\textdegree & 0\textdegree & 0 & 4 & 3 & 0.5\textdegree & 0\textdegree & 1 & 2 & 3 & 1.0\textdegree & -7\textdegree & -0.0\textdegree\\
\enddata
\end{deluxetable*}

\clearpage
\bibliography{mkat_beam_uls}{}
\bibliographystyle{aasjournal}

\end{document}